\documentclass[]{spie}  

\usepackage{amsmath,amsfonts,amssymb}
\usepackage{graphicx}
\usepackage[colorlinks=true, allcolors=blue]{hyperref}
\usepackage{siunitx}
\usepackage{placeins}
\usepackage{multicol}
\usepackage{multirow}
\usepackage{array}
\usepackage{booktabs}
\usepackage{setspace}
\usepackage{tocloft}
\usepackage{subcaption}
\usepackage{wrapfig}
\usepackage{makecell}

\title{The XOC X-ray Beamline: Probing Colder, Quieter, and Softer}

\author[a,b]{Haley R. Stueber}
\author[a]{Tanmoy Chattopadhyay}
\author[a]{Sven C. Herrmann}
\author[a]{Peter Orel}
\author[a,b]{Tsion Gebre}
\author[a,b]{Aanand Joshi}
\author[a,b,c]{Steven W. Allen}
\author[a,c]{Glenn Morris}
\author[a]{Artem Poliszczuk}
\affil[a]{Kavli Institute for Particle Astrophysics and Cosmology, Stanford University, 452 Lomita Mall, Stanford, CA 94305, USA}
\affil[b]{Department of Physics, Stanford University, 382 Via Pueblo Mall, Stanford CA 94305, USA}
\affil[c]{SLAC National Accelerator Laboratory, 2575 Sand Hill Road, Menlo Park, CA 94025, USA}

\authorinfo{Further author information: (Send correspondence to H.R. Stueber)\\H.R. Stueber: E-mail: hstueber@stanford.edu, Telephone: 1 920 252 5189}

\begin{document} 
\maketitle

\begin{abstract}
Future strategic X-ray satellite telescopes, such as the probe-class Advanced X-ray Imaging Satellite (AXIS), will require excellent soft energy response in their imaging detectors to enable maximum discovery potential. In order to characterize Charge-Coupled Device (CCD) and Single Electron Sensitive Read Output (SiSeRO) detectors in the soft X-ray region, the X-ray Astronomy and Observational Cosmology (XOC) group at Stanford has developed, assembled, and commissioned a 2.5-meter-long X-ray beamline test system. The beamline is designed to efficiently produce monoenergetic X-ray fluorescence lines in the 0.3-10 keV energy range and achieve detector temperatures as low as 173 K. We present design and simulation details of the beamline, and discuss the vacuum, cooling, and X-ray fluorescence performance achieved. As a workhorse for future detector characterization at Stanford, the XOC beamline will support detector development for a broad range of X-ray astronomy instruments.
\end{abstract}

\keywords{X-ray, CCD, SiSeRO, low noise, readout, test system, vacuum chamber, AXIS}

\section{INTRODUCTION}
\label{sec:intro}  

Providing simultaneous imaging and moderate spectroscopic capabilities for X-ray energies from $\sim 0.3-10$ keV, charge-coupled device (CCD) detectors revolutionized the field of X-ray astronomy \cite{Janesick01, Lesser15_ccd, gruner02_ccd}. 
The CCD detectors onboard the two current flagship X-ray observatories, the \emph{Chandra X-ray Observatory} and the \emph{X-ray Multi-Mirror Mission} (XMM-Newton), have enabled groundbreaking insights across a myriad of astrophysical phenomena by providing high resolution imaging of X-ray sources. Although Chandra and XMM-Newton have accomplished much throughout their lifetimes, they were deployed nearly 25 years ago and there is an ever-widening gap between the capabilities of these operating X-ray observatories and their newer infrared/optical counterparts. In order to close this gap, the 2020 Decadal Survey of Astronomy and Astrophysics \cite{Decadal23} prioritized a new great X-ray observatory flagship, as well as a nearer-term X-ray probe-class mission that would build upon the successes of Chandra and XMM-Newton, providing order of magnitude improvements in performance.
For example, among the X-ray probe-class mission candidates, the \emph{Advanced X-ray Imaging Satellite} (AXIS) would uniquely provide arcsecond spatial resolution imaging across a wide field of view with a mirror collecting area 10 times that of Chandra. These characteristics will enable transformative measurements of faint sources, providing a natural complement to optical/infrared telescopes such as \emph{Roman} and the \emph{James Webb Space Telescope} (JWST). However, for brighter sources, the narrow point-spread function (PSF) and large collecting area of AXIS increase the risk of photon pileup, impeding accurate flux and energy measurements \cite{Ballet99, McCollough05, lumb00_pileup_xmm}. Minimizing pileup and saturation effects for the small pixel, large focal plane detectors required of next generation X-ray telescopes necessitates improvements in detector technology and fast, low-noise, low-power readout methods. 

Much progress has been made in developing technology that has the potential to fulfill these requirements, including recent work done with Active Pixel Sensors (APS) such as Hybrid CMOS detectors (HCDs) \cite{HCMOS07, HCMOS17} and Depleted Field Effect Transistor (DEPFET) detectors \cite{DEPFET20}. However, current HCDs suffer from relatively high readout noise\cite{chattopadhyay18_HCDoverview}, while DEPFET X-ray imaging detectors focus on relatively large pixel sizes\cite{andricekDEPFET22}, making them less-than-ideal candidates for missions targeting high spatial resolution. X-ray CCD technology, however, can deliver the speed and noise performance required for missions like AXIS. To this end, in collaboration with MIT and MIT Lincoln Laboratory, Stanford University is developing integrated electronics to rapidly read out these detectors with high speed, low noise, and low power consumption. To operate these detectors at high speeds and characterize their spectroscopic and noise performance in a way that reflects the conditions and requirements for future missions, we needed a test system that could probe the full range of necessary X-ray energies and operating temperatures, under representative vacuum conditions. To accomplish this, we have constructed and commissioned the XOC X-ray Beamline at Stanford. In this manuscript, we detail its design and demonstrated performance, and present some initial X-ray spectra gathered using the beamline.

\section{The XOC X-ray Beamline Experimental Test System}

\begin{figure} [ht]
   \begin{center}
   \begin{tabular}{c}
   \includegraphics[height=5.8cm]{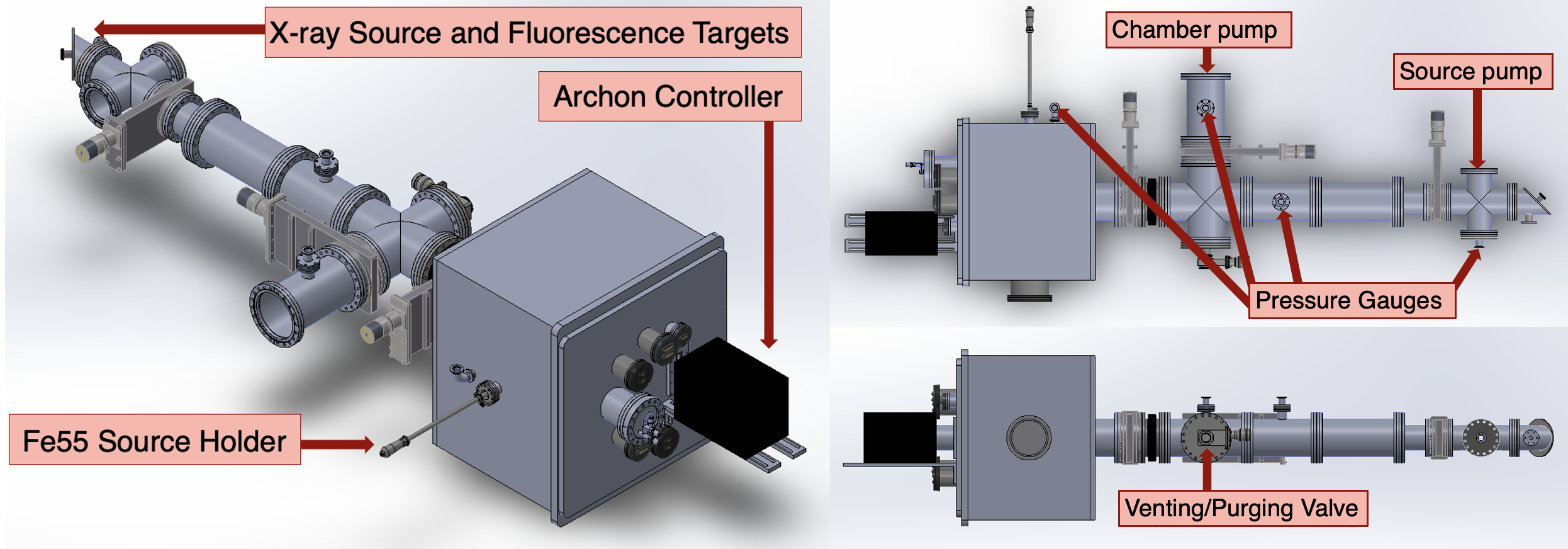}
   \end{tabular}
   \end{center}
   \caption[example] 
   {\label{fig:model} 
SolidWorks model images of the XOC X-ray Beamline. The model consists of an X-ray source and fluorescence targets, and a large chamber-side containing the detector and readout electronics, as well as an auxiliary $^{55}\mathrm{Fe}$ source. The source-side and detector chamber are separated by roughly 2 meters of standard vacuum flange components. The model view in the upper right indicates the locations of the vacuum pumps and pressure gauges on the beamline. The model view in the lower right indicates the location of the valve used for venting the beamline.}
\end{figure}

\begin{figure} [ht]
   \begin{center}
   \begin{tabular}{c}
   \includegraphics[height=8.5cm]{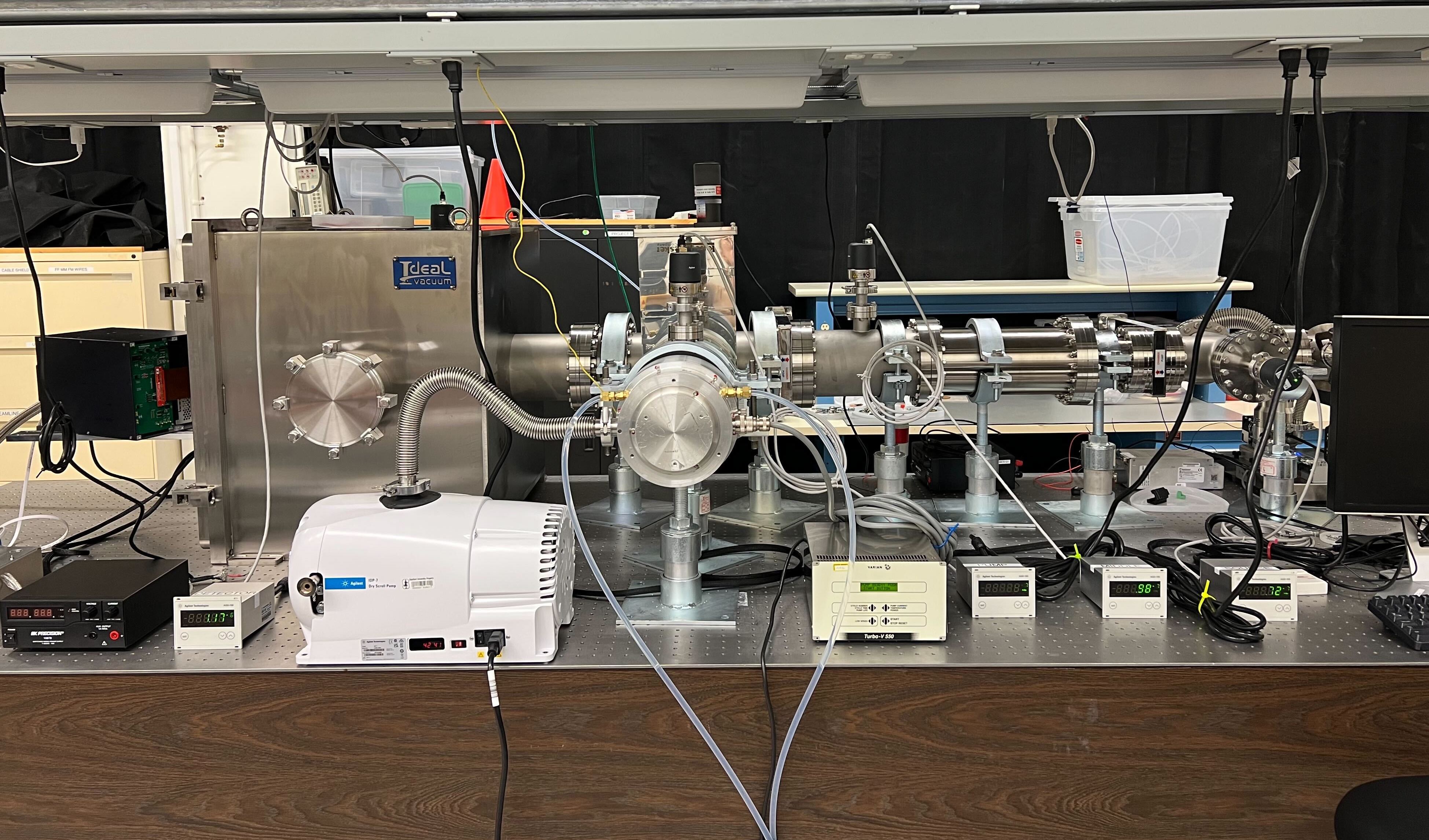}
   \end{tabular}
   \end{center}
   \caption[example] 
   {\label{fig:assembly} 
Fully assembled XOC X-ray beamline on top of an optical table. In the image, the far left side of the beamline is the detector chamber-side, while on the far right is the source-side.}
   \end{figure}

\begin{figure} [ht]
   \begin{center}
   \begin{tabular}{c}
   \includegraphics[height=6.5cm]{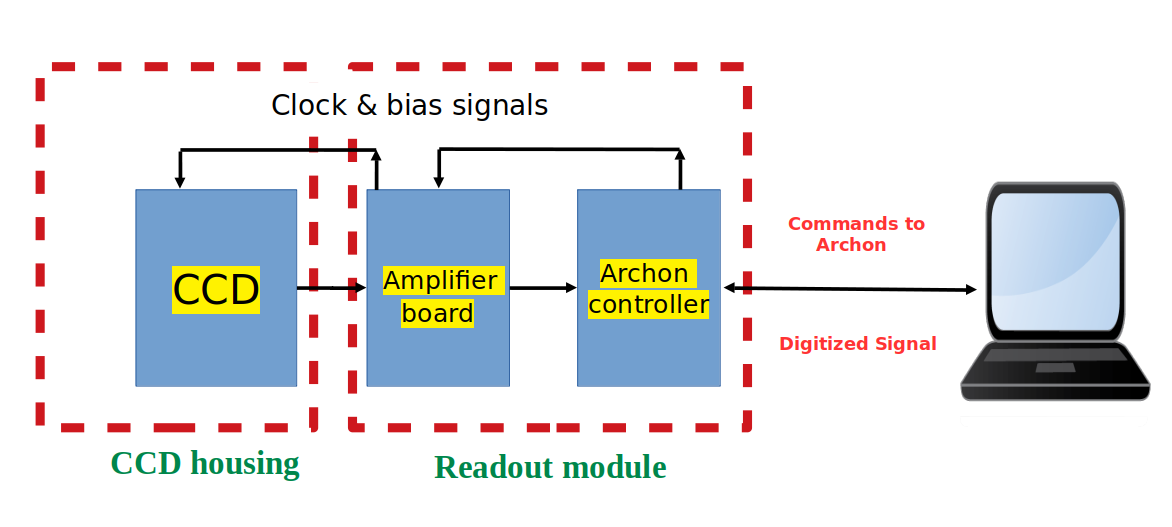}
   \end{tabular}
   \end{center}
   \caption[example] 
   {\label{fig:block} 
Basic readout process block diagram for XOC X-ray Beamline\cite{chattopadhyay20}. The readout module consists of an amplifier board and the Archon controller, which also sources the clock and bias signals.}
   \end{figure}

The XOC X-ray beamline fundamentally consists of an X-ray fluorescence source and a detector chamber, under vacuum and separated by roughly two meters of standard tubular vacuum flange components. The detector is mounted vertically to the inside door of the detector chamber, which also contains the cooling system, a retractable $^{55}\mathrm{Fe}$ radioisotope source, flux monitor, and detector readout electronics.  A SolidWorks\footnote{https://www.solidworks.com/} model for the XOC X-ray beamline is shown in Fig. \ref{fig:model}, and the fully assembled and commissioned beamline is pictured in Fig. \ref{fig:assembly}. The process for detector readout in the beamline is shown in the block diagram of Fig. \ref{fig:block}. The readout module consists of a fast, low-noise pre-amplifier board that receives the output signals from the detector and serves them to an Archon controller\footnote{http://www.sta-inc.net/archon/} outside the beamline, via a potted flex cable. The Archon both digitizes the output signal and supplies clock and bias signals to the CCD. The beamline detector chamber and source-side assemblies are discussed further in the following sections. We also discuss the cooling and system monitoring modules, and give a general description of the detectors and the readout electronics.

\subsection{Beamline Source-Side}
\label{sec:sourceside}

\begin{figure}
    \centering
    \begin{subfigure}{.328\textwidth}
    \includegraphics[width=\linewidth]{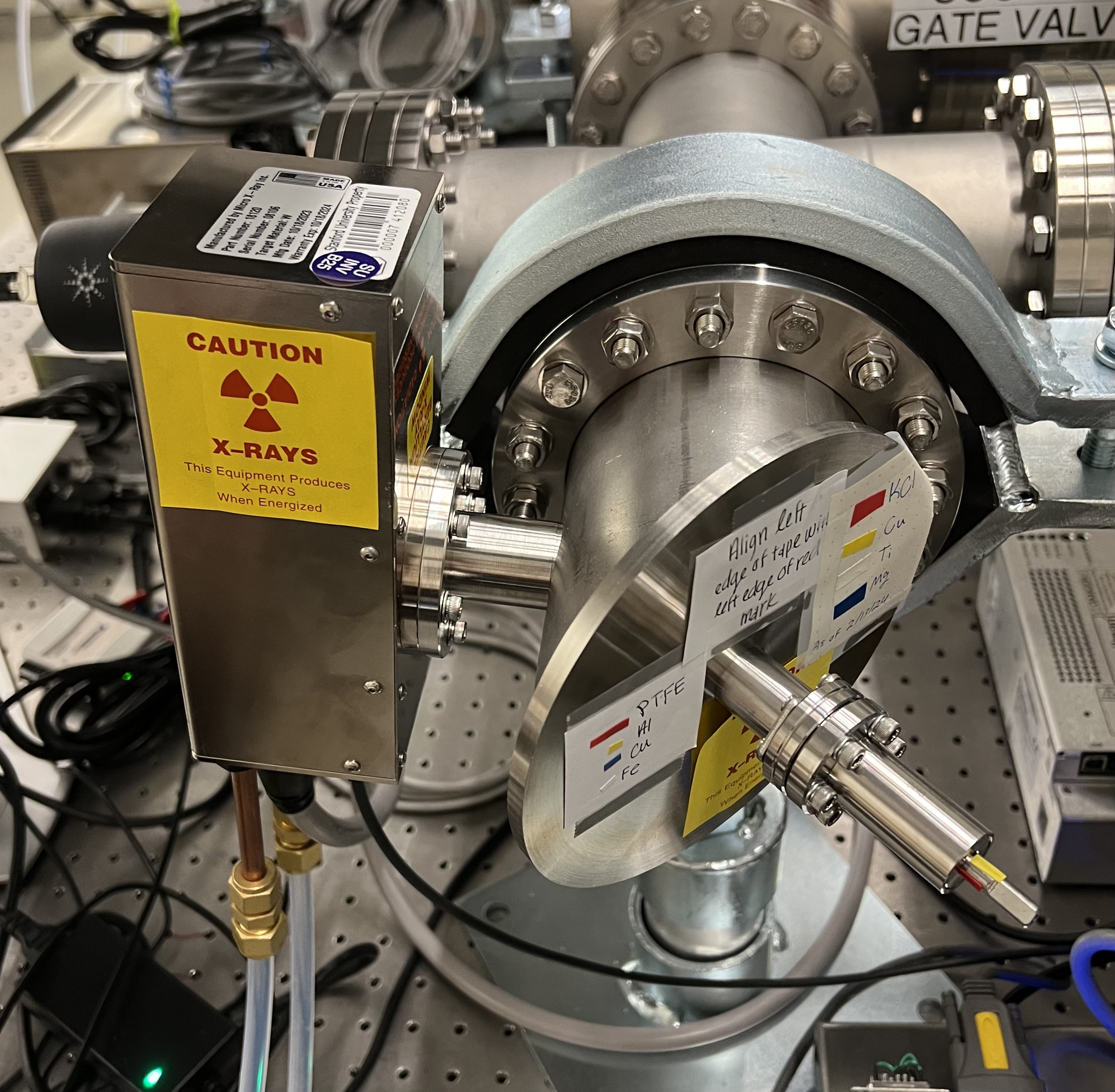}
    \caption{}
    \end{subfigure}
    \begin{subfigure}{.315\textwidth}
    \includegraphics[width=\linewidth]{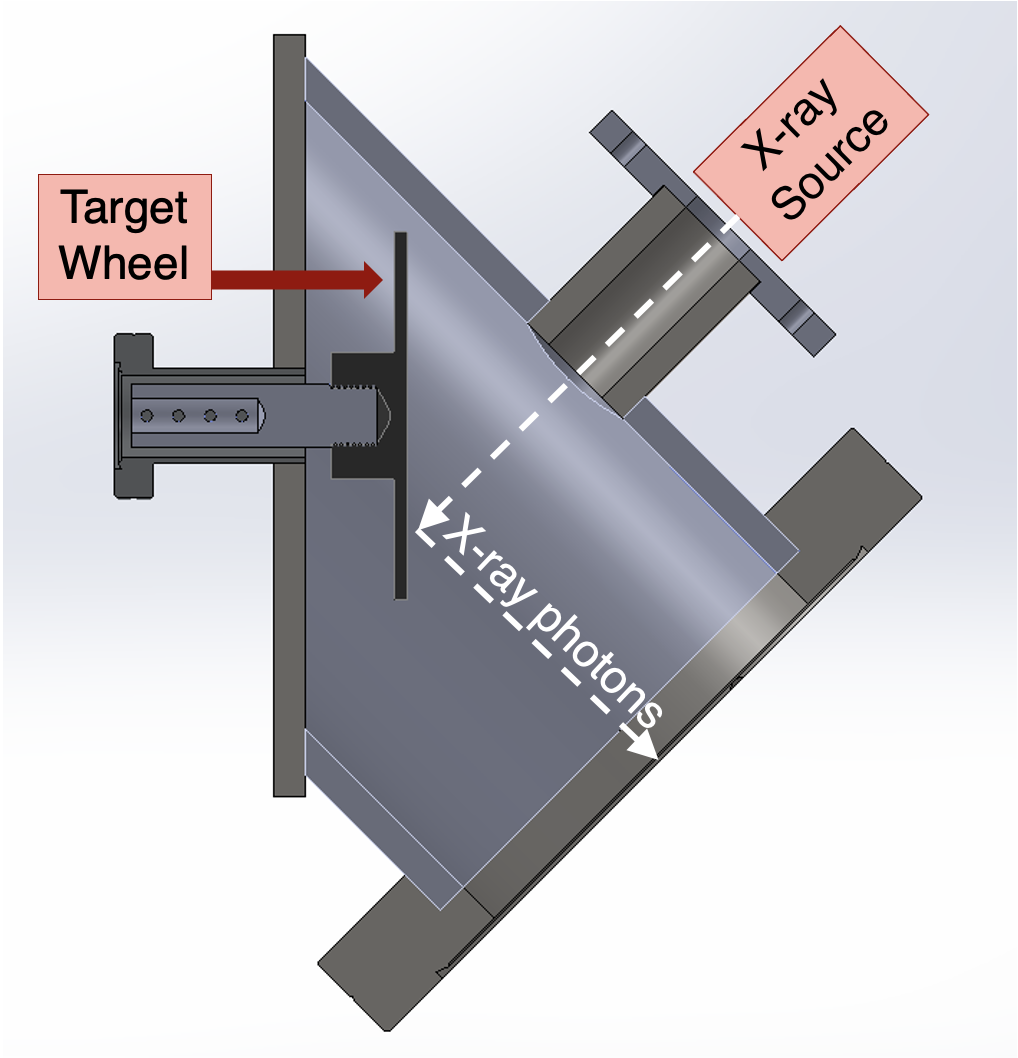}
    \caption{}
    \end{subfigure}
    \begin{subfigure}{.338\textwidth}
    \includegraphics[width=\linewidth]{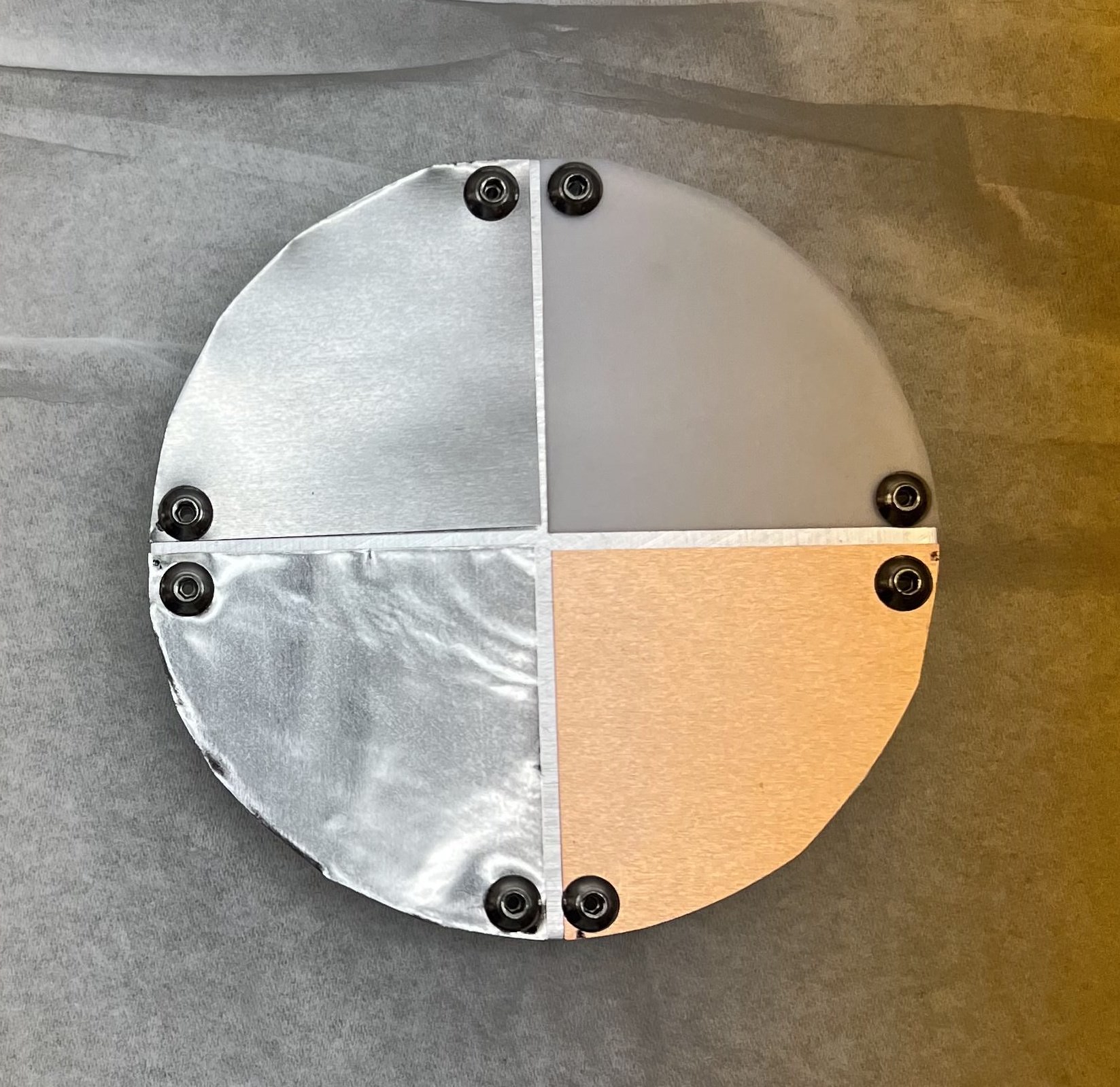}
    \caption{}
    \end{subfigure}
    \caption{(a) Beamline source end with the bremsstrahlung X-ray source attached. (b) SolidWorks model image of the source end. The white dashed arrows indicate the path of photons from the X-ray source to the target wheel, and the subsequent fluorescence photons from the target wheel into the beamline. (c) One of the target wheels with titanium (upper left), teflon (upper right), copper (lower right), and aluminum (lower left) targets attached.}
    \label{fig:source_end}
\end{figure}

The source-end of the beamline (Fig. \ref{fig:source_end}a) contains a broadband bremsstrahlung continuum X-ray source. X-rays from this source impact on a target wheel, angled at 45 degrees to the primary X-ray beam. Fluorescent X-ray photons emitted from the target travel down the length of the beamline to the detector, which is mounted on the chamber door. The process is shown in Fig. \ref{fig:source_end}b. A range of target materials are accessible on two rotatable, interchangeable target wheels, enabling the beamline to produce mono-energetic lines spanning the entire energy range of interest. The sections below detail the X-ray source specifications and performance as well as the source end model and methods.

\subsubsection{X-ray Source}
\label{sec:source}

\begin{figure} [ht]
   \begin{center}
   \begin{tabular}{c}
   \includegraphics[height=4cm]{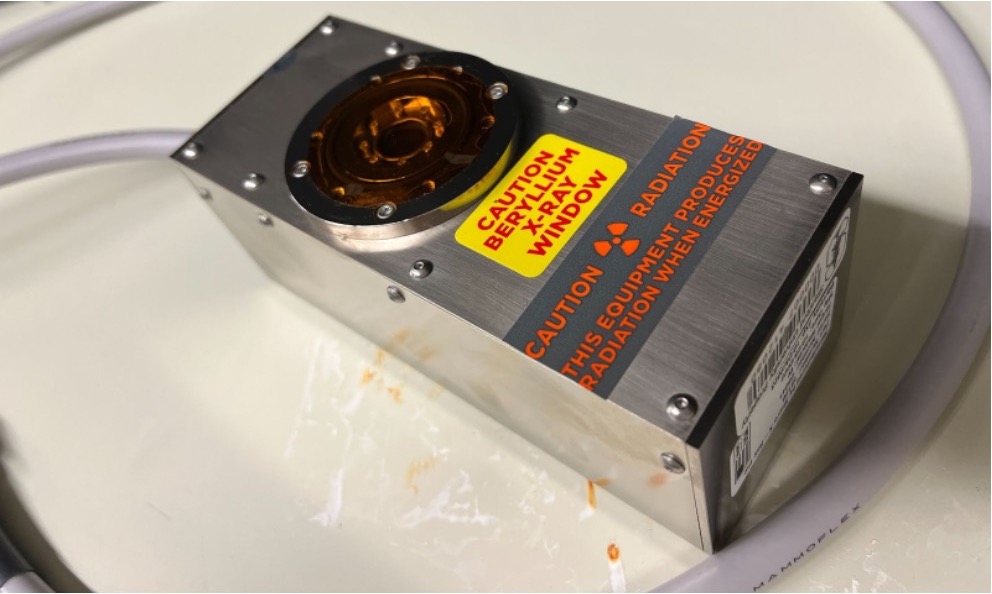}
   \end{tabular}
   \end{center}
   \caption[example] 
   {\label{fig:source} 
Micro X-ray Seeray Integrated Water-Cooled X-ray Tube X-ray source. The high voltage range is 4-60 kV, with a maximum tube current of 2.0 mA.}
\end{figure}

The primary X-ray source used in the XOC X-ray Beamline is a Micro X-ray Seeray 
Integrated Water-cooled X-ray Tube\footnote{https://microxray.com/products/seeray-water-cooled-x-ray-tube/}, which is pictured in Fig. \ref{fig:source}. This source is powered by a Spellman uXHP 50 kV 100 W high voltage power supply\footnote{https://www.spellmanhv.com/en/high-voltage-power-supplies/uXHP}, which features an analog interface connector and corresponding graphical user interface (GUI) that enables remote control and monitoring of the X-ray source tube voltage, emission current, and filament current. The power supply has been set up with external fail-safe interlocking, ensuring X-rays will not be generated unless the interlock switch is closed. 

The X-ray tube generates X-rays by accelerating electrons from a hot cathode filament towards a tungsten anode using high voltage. The rapid deceleration of the electrons at the anode stimulates bremsstrahlung X-ray emission. The upper limit of the energy of the photons emitted by the source depends on the high voltage potential, which is adjustable from 4-60 kV. The maximum anode current of the source is 2.0 mA, which limits the rate of X-ray photons emitted from the source. The source is protected by a 0.005 inch thick beryllium window, which is mostly transparent to X-rays excepting some attenuation at soft energies\cite{BeWindow}. 

\subsubsection{X-ray Fluorescence Methods}
\label{sec:XRF}

To produce mono-energetic X-ray photons across a broad range of energies, we use fluorescence methods. Bremsstrahlung photons from the X-ray source irradiate a target, mounted on a wheel within the source end, exciting electrons in the target material. As those electrons decay, they emit characteristic K-alpha and K-beta photons whose energy depends on the target material. Two rotatable and interchangeable target wheels are available, each carrying four separate targets (Fig. \ref{fig:source_end}c). The wheel is designed and located such that only one target is illuminated by the X-ray source at a time. In total, the two target wheels can produce 11 distinguishable fluorescence emission lines spanning the energy range 0.3 - 10 keV. The purities, characteristic energies, and thicknesses of the targets are summarized in Table \ref{tab:targets}.

\begin{table}[ht]
\caption{ Target Materials and their Purities, Characteristic Fluorescence Energies, Thicknesses, and Fano Limit Full Width Half Maximums} 
\label{tab:targets}
\begin{center}       
\begin{tabular}{|l|c|c|c|c|} 
\hline
\rule[-1ex]{0pt}{3.5ex}  \textbf{Target Material} & \textbf{Purity} & \textbf{Thickness} & \textbf{Characteristic Energies} & \textbf{Fano FWHM} \\
\hline
\rule[-1ex]{0pt}{3.5ex}  Teflon & N/A & 200 $\mu$m & 0.67 keV (Fluorine K$\alpha$)  &  39 eV \\
\hline
\rule[-1ex]{0pt}{3.5ex}  Magnesium & 99.9\% & 200 $\mu$m & 1.25 keV K$\alpha$ & 54 eV  \\
\hline
\rule[-1ex]{0pt}{3.5ex}  Aluminum & 99.999\% & 200 $\mu$m & 1.49 keV K$\alpha$ & 59 eV \\
\hline
\rule[-1ex]{0pt}{3.5ex}  Potassium Chloride & 99.8\% & 1000 $\mu$m & \makecell{2.62 keV (Chlorine K$\alpha$) \\ 3.31 keV (Potassium K$\alpha$)} & \makecell{78 eV \\ 87 eV} \\
\hline
\rule[-1ex]{0pt}{3.5ex}  Titanium & 99.99\% & 100 $\mu$m & \makecell{4.51 keV K$\alpha$ \\ 4.93 keV K$\beta$} & \makecell{102 eV \\ 107 eV} \\
\hline
\rule[-1ex]{0pt}{3.5ex}  Iron &  99.99\% & 125 $\mu$m & \makecell{6.40 keV K$\alpha$ \\ 7.06 keV K$\beta$} & \makecell{121 eV \\ 128 eV} \\
\hline
\rule[-1ex]{0pt}{3.5ex}  Copper &  99.99\% & 100 $\mu$m & \makecell{8.05 keV K$\alpha$ \\ 8.91 keV K$\beta$} & \makecell{136 eV \\ 143 eV}  \\
\hline
\end{tabular}
\end{center}
\end{table} 

The last column of the table lists approximate theoretical full width half maximums (FWHM) at the fano limit in silicon for each target material. These were determined assuming an average ionization energy for silicon of 3.6 eV, and the fano factor was taken to be 0.116 for silicon\cite{LOWE1997354}. Simulations of two representative target spectra are given in section \ref{sec:sims} and the measured flux and contamination for those targets are presented in section \ref{sec:detectors}.

\subsection{Detector Chamber}
\label{sec:title}

\begin{figure}
    \centering
    \begin{subfigure}{0.5\textwidth}
    \includegraphics[width=\linewidth]{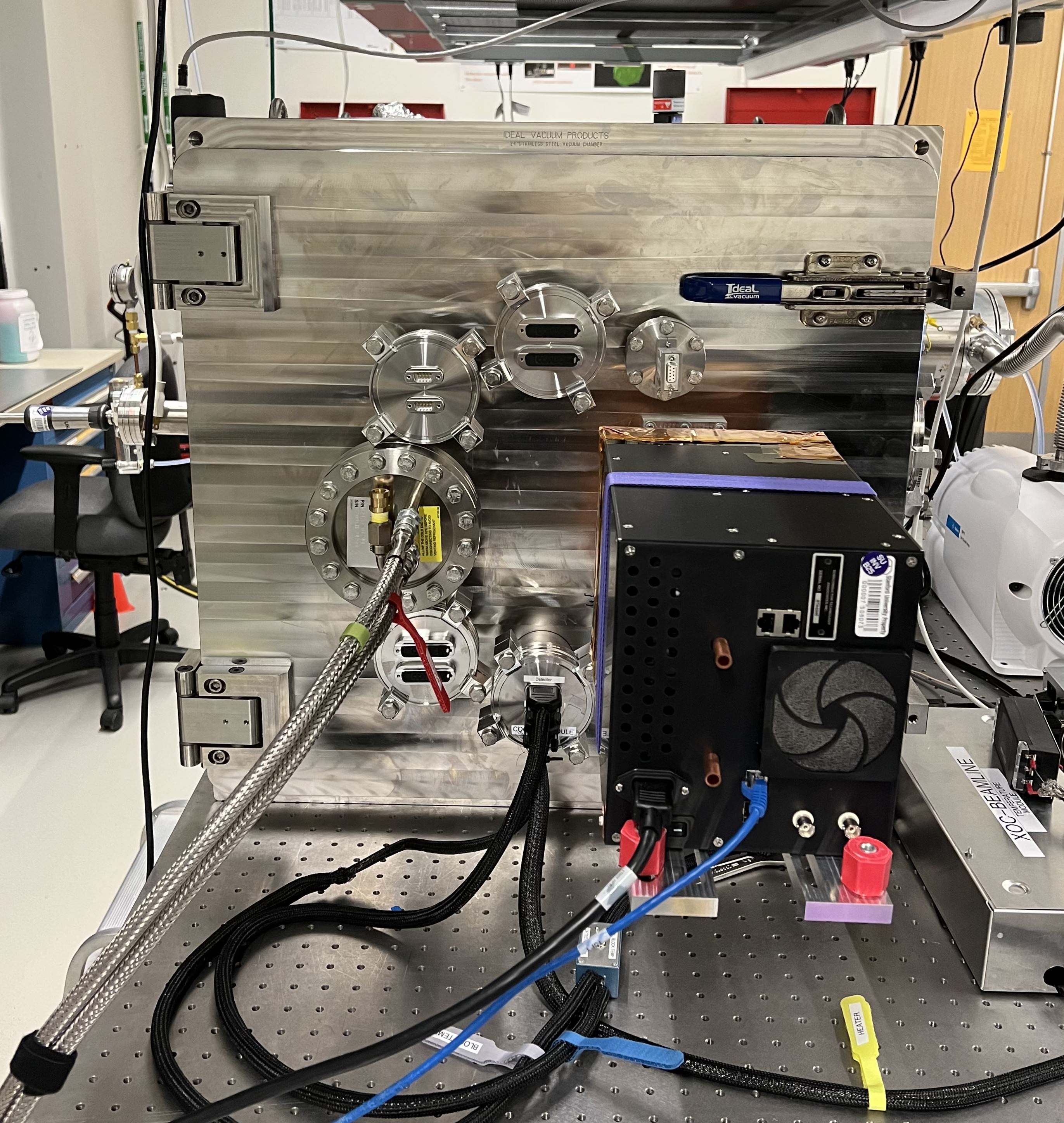}
    \caption{}
    \end{subfigure}
    \begin{subfigure}{0.397\textwidth}
    \includegraphics[width=\linewidth]{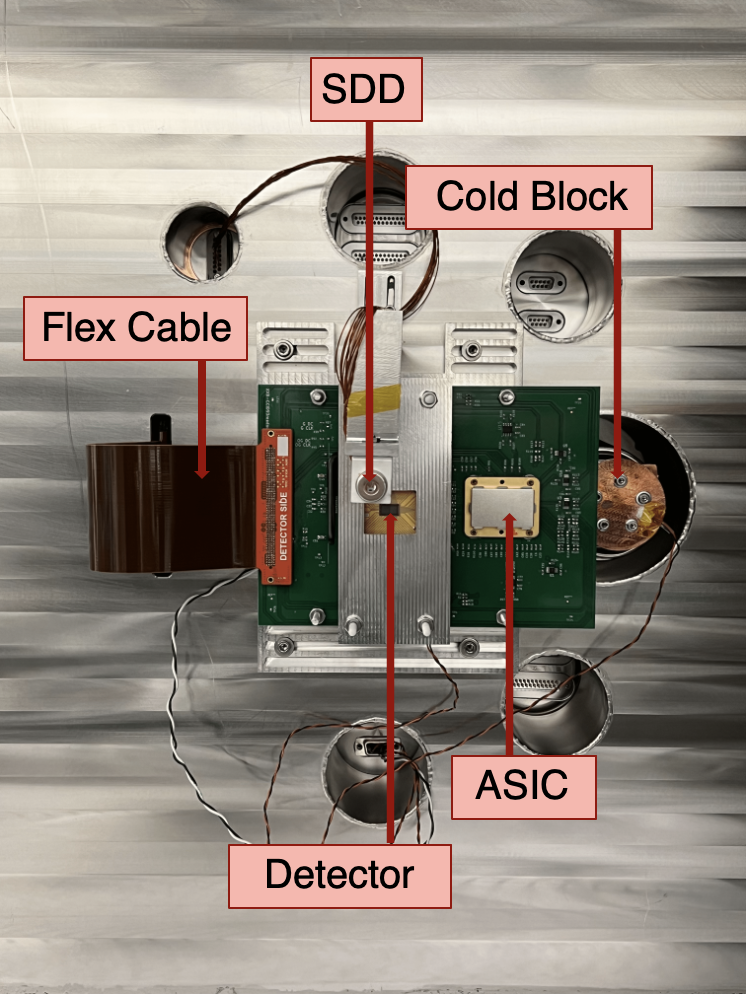}
    \caption{}
    \end{subfigure}
    \caption{(a) Detector chamber door with the Archon controller, detector temperature sensor connector flange, cryo-cooler flange, and SDD monitor connector flange. (b) Detector mounting scheme on the inside of the beamline door.}
    \label{fig:door}
\end{figure}

The main chamber of the XOC X-ray Beamline, pictured in Fig. \ref{fig:door}a, was manufactured by Ideal Vacuum Products\footnote{https://www.idealvac.com/en-us/} based on a SolidWorks model that we provided. The chamber is made of stainless steel and has dimensions of roughly 70 cm in each direction. The door of the chamber hosts standard vacuum flange components used for the vacuum pump, pressure sensor, temperature sensor, silicon drift detector (SDD), and cryocooler connections. The chamber also houses the CCD detector and its readout electronics, the cooling module, as well as an $^{55}\mathrm{Fe}$ X-ray source, which can be retracted and rotated to change the direction of radiation via a vacuum-compatible wobble stick. In this section, we outline the mechanical design of the detector chamber and detector mounting scheme, and highlight the various regulatory systems and fail-safes employed, including vacuum, venting, cooling, and flux monitoring. 

\subsubsection{Cooling and Temperature Control Module}
\begin{figure}
    \centering
    \begin{subfigure}{0.3\textwidth}
    \includegraphics[width=\linewidth]{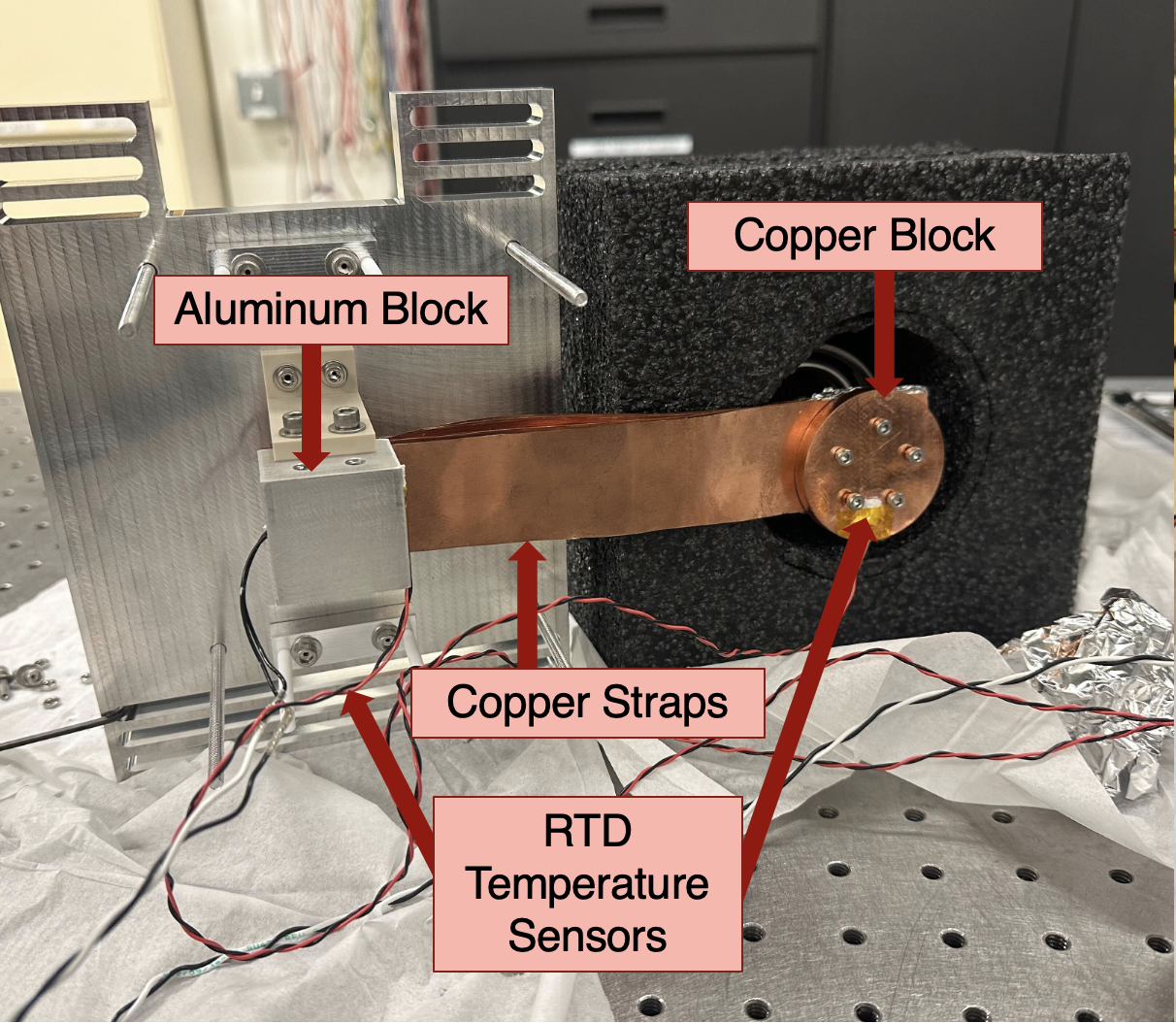}
    \caption{}
    \end{subfigure}
    \begin{subfigure}{0.35\textwidth}
    \includegraphics[width=\linewidth]{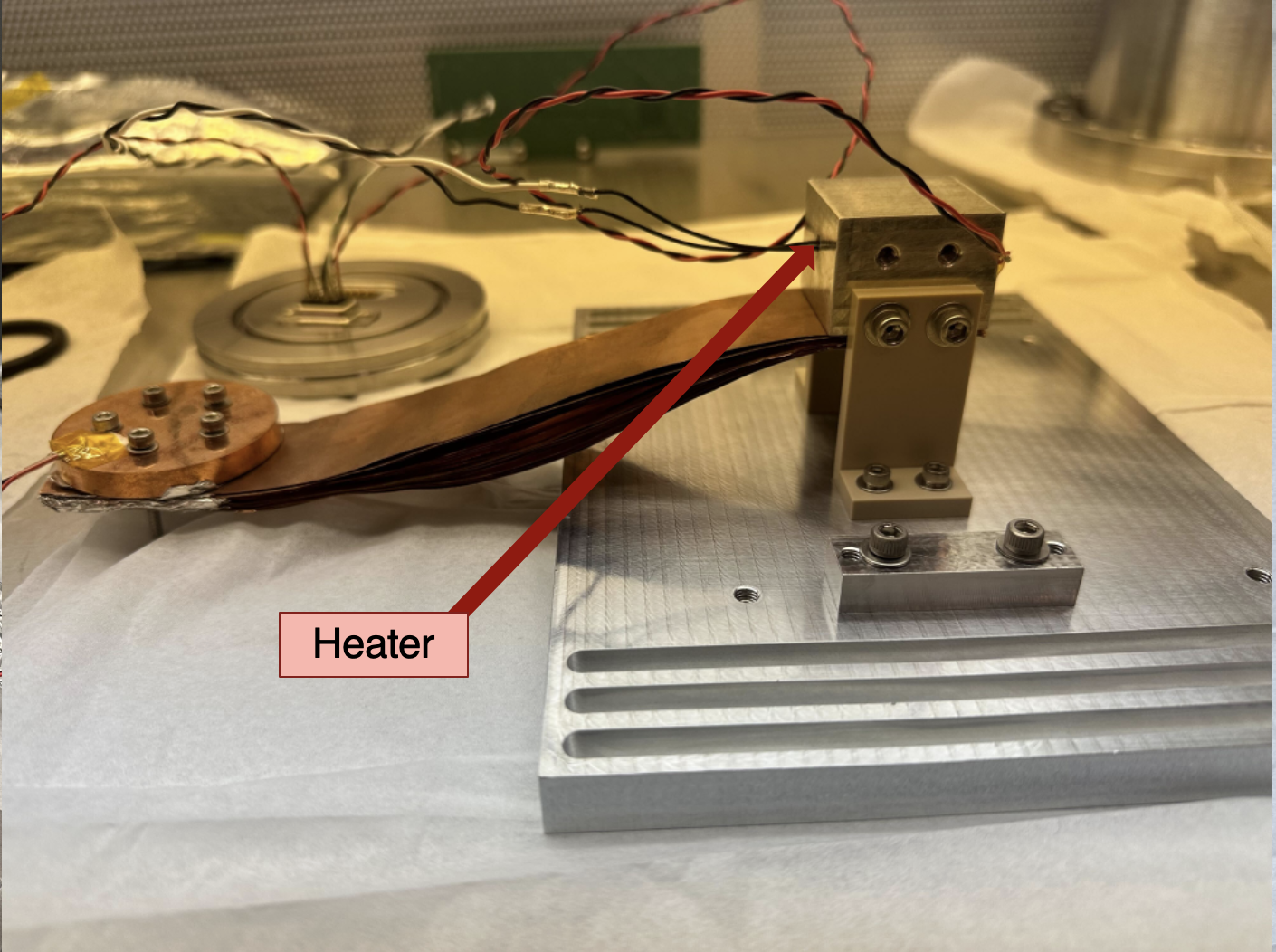}
    \caption{}
    \end{subfigure}
     \begin{subfigure}{0.254\textwidth}
    \includegraphics[width=\linewidth]{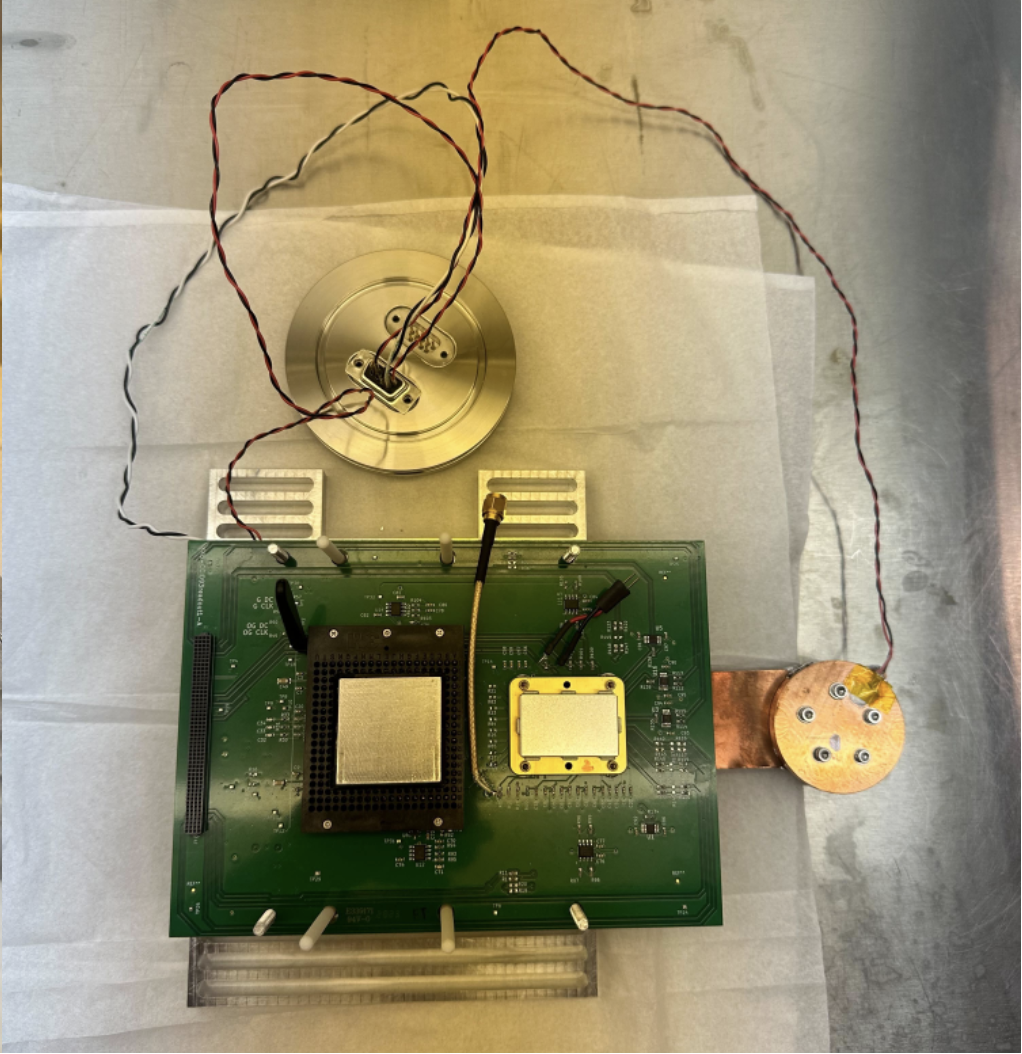}
    \caption{}
    \end{subfigure}
    \caption{(a) The detector cooling system with the aluminum cold block, copper straps, and copper block that thermally couples the system to the cold head of cryo-cooler. Resistance temperature detector (RTD) sensors are connected to the aluminum block and copper plate. (b) Heater connection inside of aluminum block. (c) Cooling system with readout PCB mounted and RTD temperature sensors connected to chamber door flange. An additional RTD sensor is soldered to the detector socket pins on the PCB.}
    \label{fig:tempstuff}
\end{figure}

A vital aspect of the beamline is its ability to efficiently cool the detector and maintain stable temperatures during testing. Efficient cooling is achieved with an Edwards Polycold PCC Compact Cryo-cooler\footnote{https://www.edwardsvacuum.com/en-us/our-markets/semiconductor-and-electronics/cryogenics/cryochillers/pcc-compact-coolers} which, by itself, is able to achieve temperatures as low as 70 K. The cold-end of the cooler is mounted to the detector chamber door, and it is thermally coupled to an aluminum block atop which the detector housing sits. This thermal connection is achieved by clamping a stack of thin copper film straps and indium foils to the cold head with a copper block. The other end of the straps is clamped to the bottom of the aluminum block under the detector, which is supported using thermally insulated peek plastic standoffs. A layer of Tflex-500 thermally conductive gap filler adhesive padding\footnote{https://www.laird.com/sites/default/files/2019-10/Tflex\%20500\%20Series.pdf}
is sandwiched between the aluminum block and the back of the detector package to ensure uniform thermal contact. This setup and its components are shown in Fig. \ref{fig:tempstuff}. 

It is important during operation that the CCD parameters do not have any temperature dependent fluctuations. We use RTD sensors coupled with a proportional-integral-derivative (PID) temperature control loop to monitor and stabilize the detector temperature\cite{WangPID2018}. There are two RTDs in the system; one is adhered via a thermally conductive but electrically insulating epoxy to the copper plate on the cold-head, and the other is soldered onto pins of the detector socket, which is mounted to the readout preamplifier board. Such a configuration makes it possible to monitor both the detector and the cryo-cooler cold-end temperatures. These 2-wire PT 1000 Tewa Sensor LLC RTDs\footnote{https://www.digikey.com/en/products/detail/tewa-sensors-llc/TT-PT-1000B-2050-11-AUNI/9817197} are wired to their respective Adafruit MAX31865 RTD temperature sensor amplifier boards\footnote{https://www.adafruit.com/product/3648}, which are operated by a Raspberry Pi\footnote[1]{https://www.raspberrypi.com/}. As such, the Raspberry Pi both controls the bias and the clocking conditions of the temperature sensors and runs the PID temperature control loop. The PID regulates the detector temperature by reading out the adjacent RTD sensor and using it as the input for the PID feedback loop along with the reference temperature setpoint given by the user. The PID actuator is a Birk Manufacturing cartridge heater\footnote[2]{https://www.digikey.com/en/products/detail/birk-manufacturing/DC1101/15790977} embedded in the aluminum block. With these cooling methods we are able to achieve stable operation down to 173 K with a temperature stability of roughly 0.1 K.

\subsubsection{System Monitoring and Fail-Safes}
\label{sec:monitor}
To ensure safe and reliable operation of the beamline, we have implemented a variety of fail-safes and monitoring systems. Namely, these include a pressure monitoring and alarm system, venting and purging with clean dry air and a pressure relief valve, back-up power supplies for the vacuum pumps, and the X-ray source interlocking fail-safe outlined in section \ref{sec:source}. Two vacuum pumps are used to pump out the beamline to pressures as low as 5e-8 mbar. An Agilent IDP-7 Scroll Pump\footnote[3]{https://www.agilent.com/store/productDetail.jsp?catalogId=X3807-64000\&catId=SubCat2ECS\_1465486} and Varian TV 551 Navigator Turbo Pump\footnote[4]{https://www.ajvs.com/product{\textunderscore}info.php?products\_id=17586\&1808} are attached to the chamber-side of the beamline, and an Agilent TPS-flexy Turbo Pumping System\footnote[5]{https://www.agilent.com/store/productDetail.jsp?catalogId=X1699-64087\&catId=SubCat2ECS\_1466018} pumps out the source-side. Gate valves at different positions throughout the beamline make it possible to bring the chamber to atmosphere while the main body and source end of the beamline can remain under vacuum. A venting valve positioned near the detector chamber (Fig. \ref{fig:model}) can be opened to purge the chamber with clean dry air. For general safety and the safety of the detector and SDD monitor, we have a pressure relief valve which will open in the event of pressures exceeding 1.2 atmospheres in the chamber during purging. For monitoring the pressure, four pressure gauges are mounted at different points in the beamline, indicated in the aerial view of the SolidWorks beamline model in Fig. \ref{fig:model}. The first gauge monitors the pressure in the detector chamber, the second monitors the pressure near the main chamber pump, the third monitors pressure in the middle of the beamline, and the last monitors the pressure near in source end of the beamline. The pressure monitors used are Agilent FRG-700 Full Range Pirani Inverted Magnetron Gauges\footnote[6]{https://www.agilent.com/store/productDetail.jsp?catalogId=FRG700KF25\&catId=SubCat2ECS\_1465644}. The pressure in each sector of the beamline is recorded continuously via a Python program, and if it rises above a critical pressure of 5e-5 mbar, Slack\footnote[7]{https://slack.com/} alerts are sent to all lab members every five minutes until sub-critical pressure is restored. The vacuum pumps are attached to uninterruptible power supplies (UPS) to ensure vacuum is maintained even in the event of a power outage. In this way, it is possible to prevent a sudden loss of vacuum, which could damage the source and lead to condensation buildup on the detector while cold.

\subsubsection{Silicon Drift Detector Flux Monitor}
\label{sec:sdd}
\begin{figure} [ht]
   \begin{center}
   \begin{tabular}{c}
   \includegraphics[height=7cm]{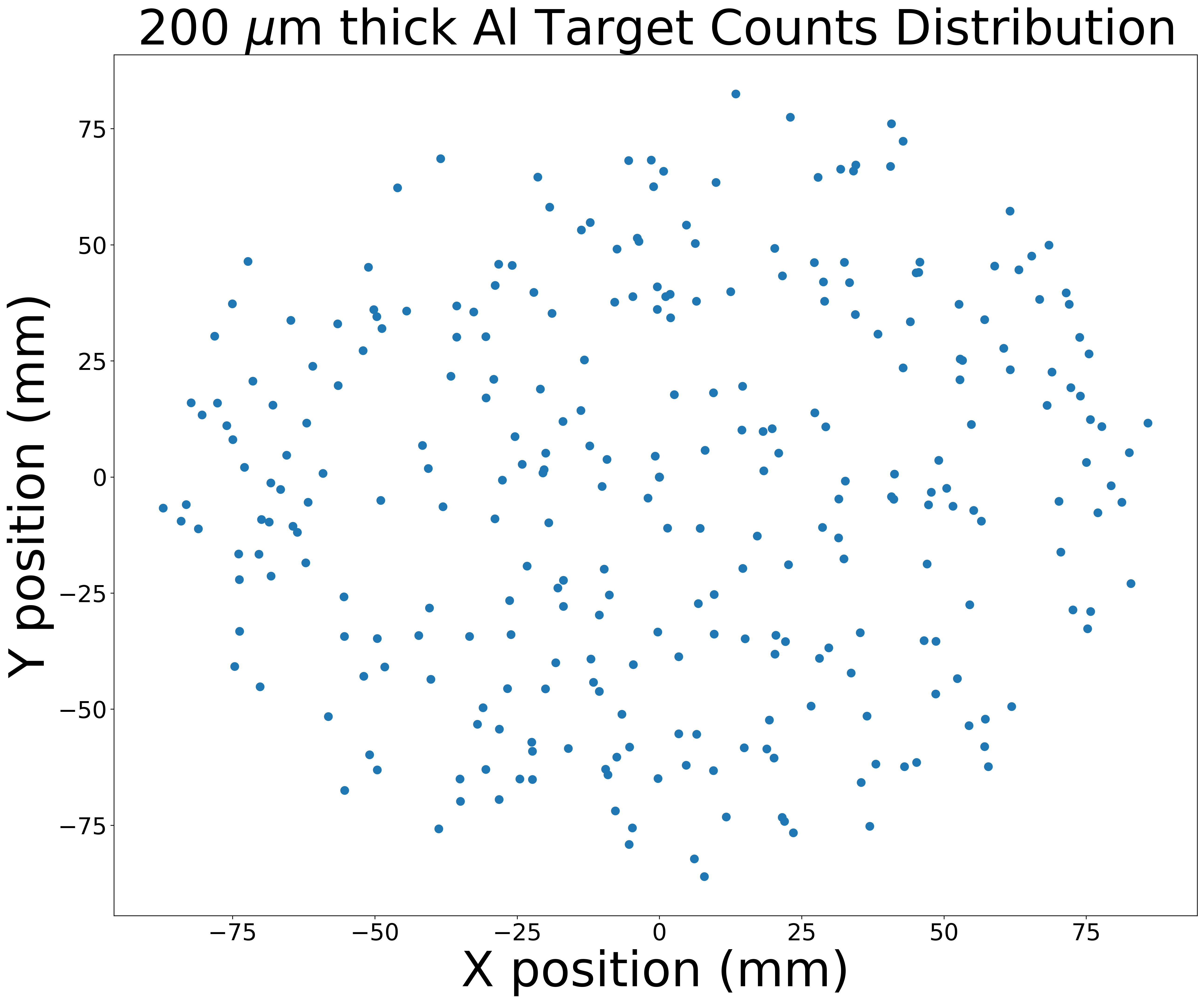}
   \end{tabular}
   \end{center}
   \caption[example] 
   {\label{fig:photdistrib} 
Simulated photon counts distribution on beamline detector chamber door.}
\end{figure}

To characterize detector performance, it is essential to know line fluxes of fluorescent photons within the detector chamber. To make these measurements, a silicon drift detector (SDD) is mounted on the chamber door, near the plane of the detector, as pictured in Fig. \ref{fig:door}b. Particle interaction simulations that were carried out for the beamline (discussed in \ref{sec:sims}) suggest that the photon distribution on the detector chamber door appears uniform (Fig. \ref{fig:photdistrib}). Therefore, by being in the field of view, the SDD provides accurate and simultaneous measurements of incident flux and can be used for fast and easy detector quantum efficiency estimates. SDD detectors generally consist of a window, a fully depleted silicon bulk, and a collection anode, the small size of the which leads to a small device capacitance. Because of this, these devices typically demonstrate very low noise and high resolution at high count rates\cite{SDD}. The beamline SDD model is an Amptek 25 $mm^{2}$ Fast Silicon Drift Detector\footnote[8]{https://www.amptek.com/products/x-ray-detectors/fastsdd-x-ray-detectors-for-xrf-eds/fastsdd-silicon-drift-detector} with a C2 (silicon nitride) X-ray window, which enables detection of photon energies down to 0.3 keV. The SDD comes with a built-in thermo-electric cooler (TEC) that cools the detector to 210K. The SDD is heat sunk to the detector chamber door. While the CCD is collecting data, the flux monitor continuously builds a live spectrum. This spectrum is extremely useful for aligning the targets in the source beam, making it easy to see the relative change in monochromatic line flux as the target wheel is being rotated. In this way, the SDD is also used to ensure that only one target is fully illuminated at a time.

\subsection{CCID-93 Detector and Readout Module}
\label{sec:det&readout}

In this section, we give a brief overview of the detectors currently being tested in the beamline together with their corresponding readout electronics. Details of the detectors and readout electronics can be found in [\citenum{bautz18}], [\citenum{bautz19}], [\citenum{bautz20}], [\citenum{bautz22}],
[\citenum{tanmoyJATIS22}], [\citenum{herrmann20_mcrc}], and [\citenum{porelMCRCspie2022}], while latest updates are given in proceedings [\citenum{tanmoyspie2024}], [\citenum{bevspie2024}], [\citenum{svenspie2024}], and [\citenum{porelMCRCspie2024}]. 

The CCID-93 detectors fabricated by our collaborators at MIT Lincoln Laboratory (MIT-LL) have two variants of on-chip output amplifier stages:

\begin{itemize}
    \item A source follower p-channel junction field effect transistor (p-JFET)\cite{bautz18,bautz19,bautz20,bautz22,tanmoyJATIS22}. This is a high impedance, voltage based readout method.
    \item A Single Electron Sensitive ReadOut (SiSeRO) sensor consisting of a p-channel metal-oxide semiconductor field effect transistory (p-MOSFET), which is based on a drain current readout technique\cite{ sisero2021, sisero2022, sisero2023}.
\end{itemize}  

\begin{figure} [ht]
   \begin{center}
   \begin{tabular}{c}
   \includegraphics[height=7cm]{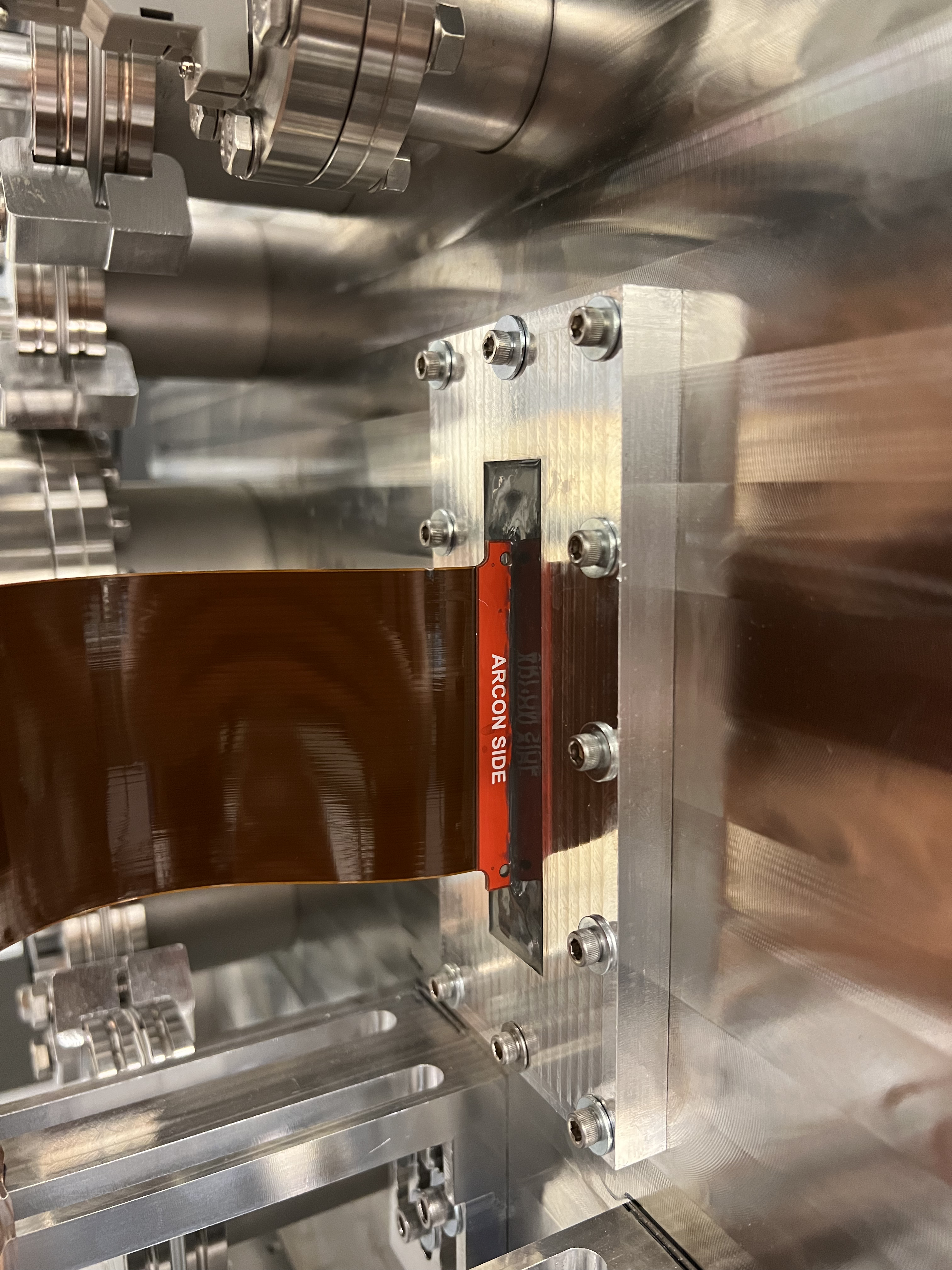}
   \end{tabular}
   \end{center}
   \caption[example] 
   {\label{fig:flex} 
200-pin flex cable potted in custom flange on detector chamber door.}
\end{figure}

\noindent The detector output signal is amplified in a preamplifier board, designed specifically to accommodate both the p-JFET and the SiSeRO output stages. The preamplifier board has the option of using either a discrete electronics implementation or an application specific integrated circuit (ASIC) for the readout. The ASIC used is the multi-channel readout chip (MCRC)\cite{herrmann20_mcrc, MCRC22,porelMCRCspie2022} developed by Stanford University. Its core advantages include a small physical footprint and low-power consumption, while providing fast, low noise readout. The preamplifier board output signal is transmitted from inside the detector chamber through a 200-pin flex cable potted in a custom-made flange (Fig. \ref{fig:flex}) to a Semiconductor Technology Associates, Inc Archon controller. The Archon is an FPGA-based instrument that provides control signals like clocks and biases to the detector and simultaneously digitizes the signal of interest. In addition, the Archon controller features an internal digital signal processing pipeline that extracts the relevant information from the acquired detector waveform to reconstruct a 2-dimensional image. A more detailed description of the Archon controller modules is given in [\citenum{chattopadhyay20}].

\section{GEANT4 Simulations of Beamline Performance}
\label{sec:simulations}

Before fabrication, to optimize the performance of the XOC beamline, including line fluxes, spectral contamination and radiation leakage, we carried out Monte Carlo simulations of the particle interactions that would take place during operation of the source. To model these complex particle-matter interactions, the open source {\ttfamily GEANT4} toolkit\footnote[9]{https://geant4.web.cern.ch/}\cite{AGOSTINELLI2003} was used, which includes a comprehensive set of physics processes for electromagnetic, strong, and weak interactions of particles in matter over a broad energy range. {\ttfamily GEANT4} contains an extensive database of materials to assign to a number of pre-defined geometries as well as the option to import computer aided design (CAD) models. In the sections below, we describe the model geometries and material properties assigned in the {\ttfamily GEANT4} model to each of the beamline components, and define the interaction physics included in the model. We also present simulation results for some representative targets.

\subsection{Geometries and Materials}
\label{sec:g4model}

In order to build our model geometries and define their material compositions, we utilize a modified version of the {\ttfamily GEANT4} detector construction class {\ttfamily G4VUserDetectorConstruction}. Geometries can be defined in {\ttfamily GEANT4} either by using built-in shape functions such as {\ttfamily G4Box} or {\ttfamily G4Sphere}, or by importing CAD models. For example, the beamline SolidWorks models for the detector chamber, main body, and source end were imported as .stl files into {\ttfamily GEANT4} using the {\ttfamily CADmesh2} library\cite{poole11}. These components were assumed to be made of 304L stainless steel (69\% iron, 18.5\% chromium, 9\% nickel, 2\% manganese, and 1\% silicon with less than 1\% of carbon, phosphorous, sulfur, and nitrogen). The chemical compositions defined in the simulations for 304L stainless steel and all other materials are detailed in Appendix \ref{sec:materials}.

The targets and target wheel were defined as {\ttfamily G4Box} geometries. The targets were assumed to be pure with the exception of the teflon target, which was defined as 24\% carbon and 76\% fluorine. The material of the wheel itself was defined as an aluminum-6061 alloy. The X-ray source contains a beryllium window which causes some attenuation of soft X-rays, so this window was also included in the simulations. The detector is defined as an active volume made of silicon, which means that during simulation, particle interaction types and positions as well as energy depositions in the detector volume are recorded. Dimensions of the detector box geometry were given as 200mm x 200mm with thickness 0.1mm. To reduce simulation times and to accumulate enough detector photon counts to produce a simulated spectrum, the detector area used in simulation was scaled up by a factor of about 2500 from that of a real CCID-93 detector. Since the actual expected number of counts scales directly with detector area, the actual expected number of detected photons would be 2500 times less than is achieved in simulation. 

For computational simplicity, it was assumed that smaller-scale elements such as screws, rods and wires would not have a large impact on the simulation results and were therefore not included in the model. However, elements in direct contact with or near to the detector were defined as simplified block geometries with their corresponding material properties in order to estimate their contributions to spectral contamination. Such components included the copper straps and block which thermally couple the detector to the cryo-cooler head, the G10 plastic standoffs which support the cold block under the detector, the PCB board on which the detector is mounted, and the ceramic detector housing itself.

\subsection{Simulation Physics}

For the particle interaction physics used in our simulations, we have user defined physics lists from the {\ttfamily GEANT4} {\ttfamily G4VUserPhysicsList} class. The physics list defines processes involving X-ray photons and secondary electrons generated from X-ray interactions with the beamline geometries. The processes we've included in our simulations are listed in Appendix \ref{sec:g4physics}.

\begin{figure} [ht]
   \begin{center}
   \begin{tabular}{c}
   \includegraphics[height=7cm]{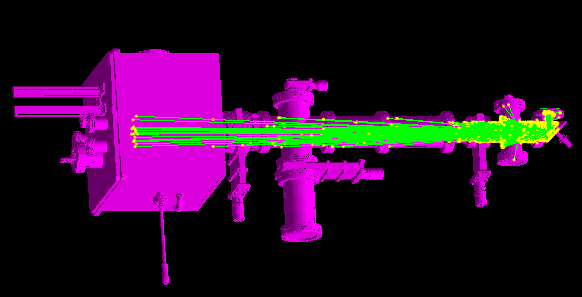}
   \end{tabular}
   \end{center}
   \caption[example] 
   {\label{fig:g4model} 
Visualization of GEANT4 simulation of particle interactions inside the beamline after 400,000 events have taken place. Green tracks indicate photons and yellow dots indicate interaction locations.}
\end{figure}

During simulation, the {\ttfamily G4GeneralParticleSource} class is configured to generate source photons at a given position in the world volume. In this configuration, the photon source was defined as a circular plane with a momentum vector which directed X-rays into the beamline at 45 degrees from the target wheel. To emulate the photons which would be generated by the bremsstrahlung X-ray source in the beamline setup, the simulated source photons were allowed to take on energies between 1 and 20 keV according to a power law distribution with an exponent of -1. A visualization of the photon paths and interactions inside the beamline during simulation can be seen in Fig. \ref{fig:g4model}. From this visualization, we note negligible leakage of photons from the beamline.

\subsection{Simulation Results for Fluorescent Targets}
\label{sec:sims}

\begin{figure}
    \centering
    \begin{subfigure}{.45\textwidth}
    \includegraphics[width=\linewidth]{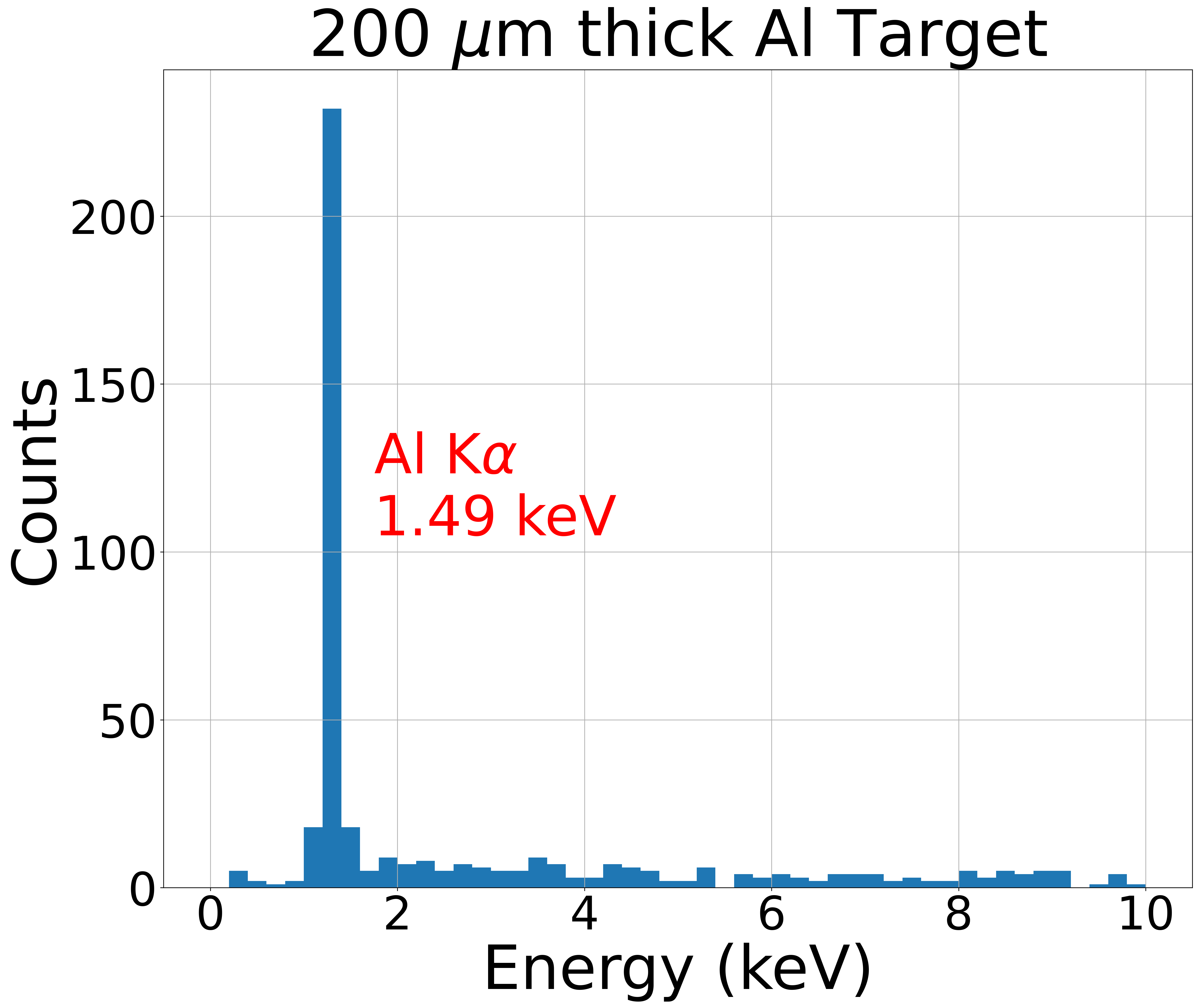}
    \caption{}
    \end{subfigure}
    \begin{subfigure}{.462\textwidth}
    \includegraphics[width=\linewidth]{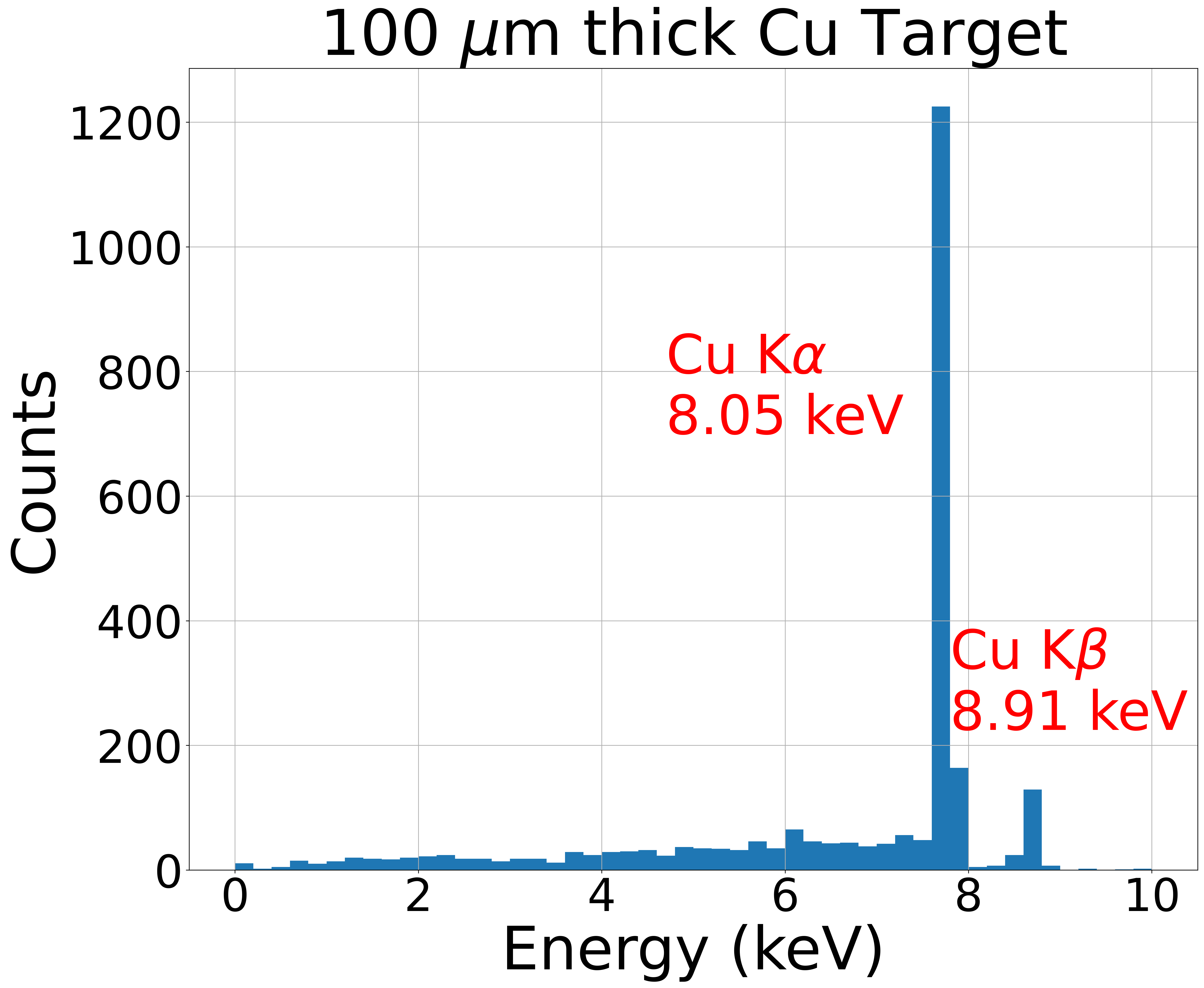}
    \caption{}
    \end{subfigure}
    \caption{Simulated GEANT4 spectra for (a) A 200 $\mu$m thick pure aluminum target (b) a 100 $\mu$m thick pure copper target.}
    \label{fig:sims}
\end{figure}

Simulations for each of the seven planned fluorescent target materials were run for 100 million source photons each. The histograms in Fig. \ref{fig:sims}a and b present the number of events per 0.2 keV bin of energy which interacted with the detector volume for the aluminum and copper targets, respectively. The characteristic K$\alpha$ and K$\beta$ lines for each target are easily identifiable, with minimal contamination at other energies observed. Based on the results of these simulations, we felt confident that the beamline would be able to produce strong, mono-energetic lines across a broad range of X-ray energies with minimal radiation leakage.

\section{Characterization of MIT-LL CCD X-ray Detectors}
\label{sec:detectors}

\begin{figure}
    \centering
    \begin{subfigure}{.48\textwidth}
    \includegraphics[width=\linewidth]{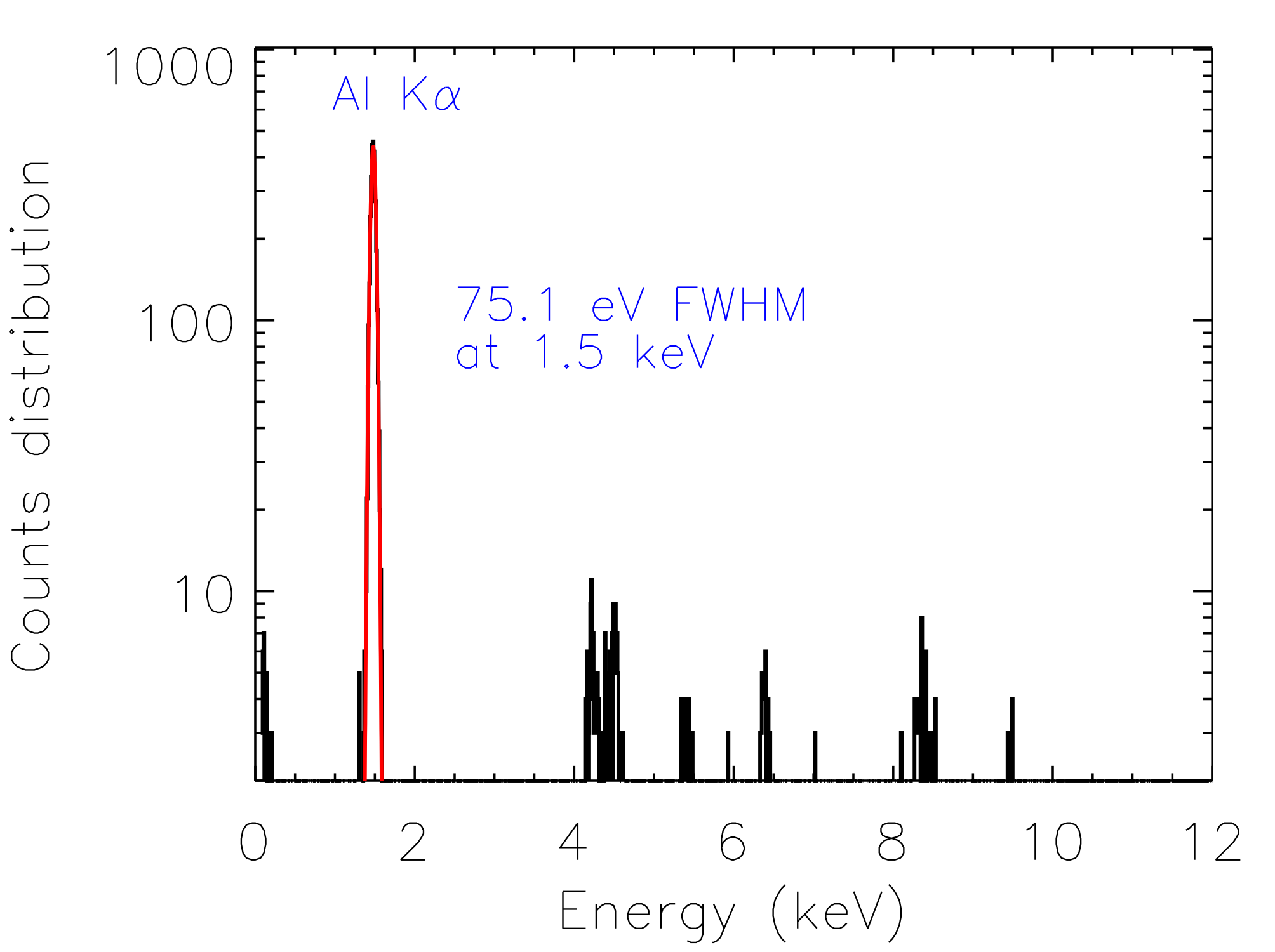}
    \caption{}
    \end{subfigure}
    \begin{subfigure}{.48\textwidth}
    \includegraphics[width=\linewidth]{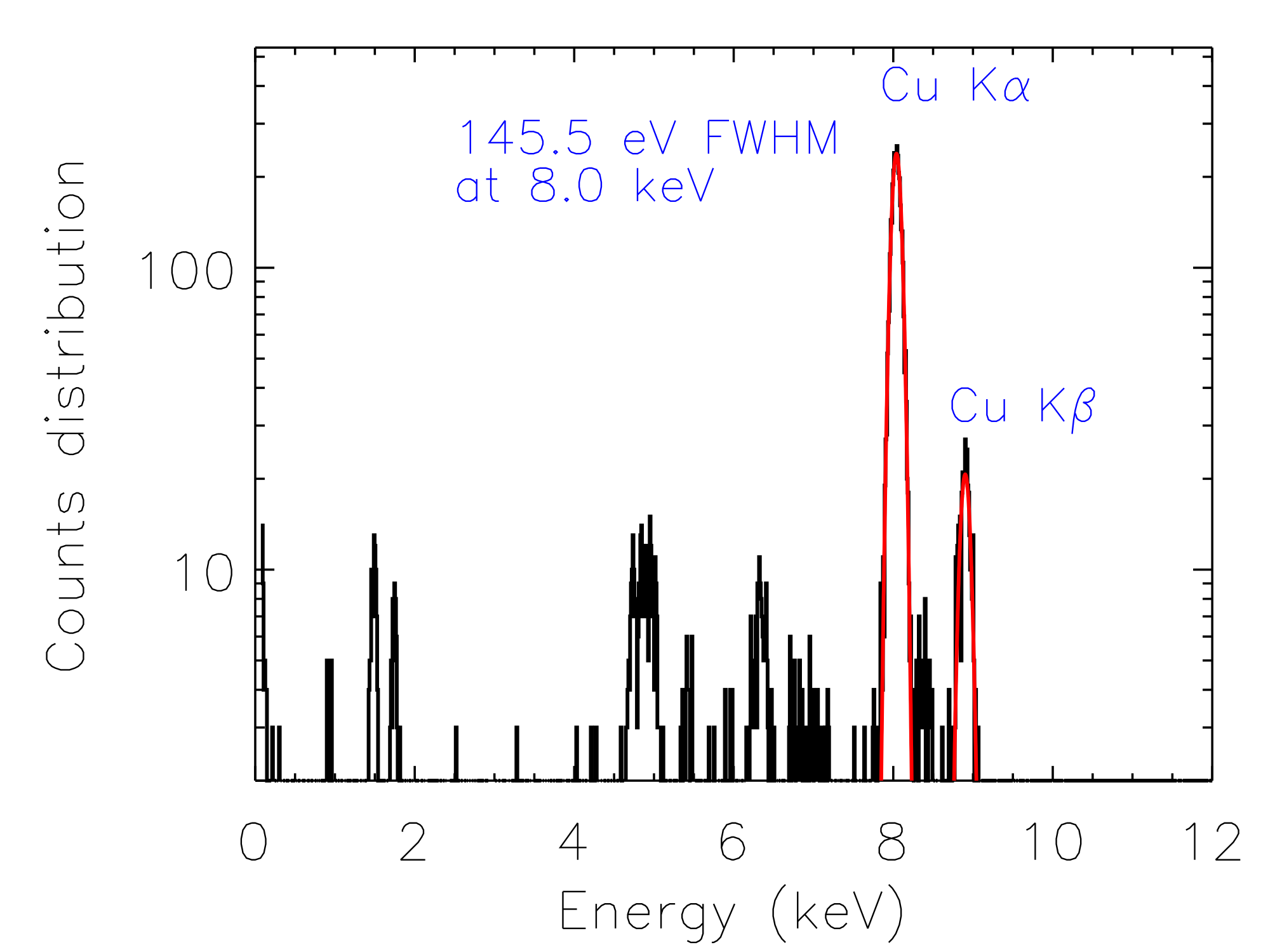}
    \caption{}
    \end{subfigure}
    \caption{CCD spectra for (a) A 200 $\mu$m thick pure aluminum target (b) a 100 $\mu$m thick pure copper target.}
    \label{fig:ccd_spectra}
\end{figure}

After the beamline had been fully assembled and commissioned, a CCID-93 detector was mounted inside. Cooled down to 193 K, the detector recorded spectra for all seven fluorescent targets. Presented here are two representative spectra for a relatively low energy line and a high energy line, respectively. The Gaussian fit to the spectrum for the Aluminum target shown in Fig. \ref{fig:ccd_spectra}a indicates a full width half maximum (FWHM) of 75.1 eV at 1.49 keV. This corresponds to a resolution $R= E/\Delta E$ of $\sim 20$. For the copper target spectrum shown in Fig. \ref{fig:ccd_spectra}b, we obtain an FWHM of 145.5 eV at the 8.05 keV K$\alpha$ energy, resulting in a resolution of $\sim 56$. In comparison to the values given in Table \ref{tab:targets}, the resolution of the aluminum line is within about $\sim 25$ \% of the fano limit, while copper is within $\sim 5$ \%.

\begin{figure} [ht]
   \begin{center}
   \begin{tabular}{c}
   \includegraphics[height=8cm]{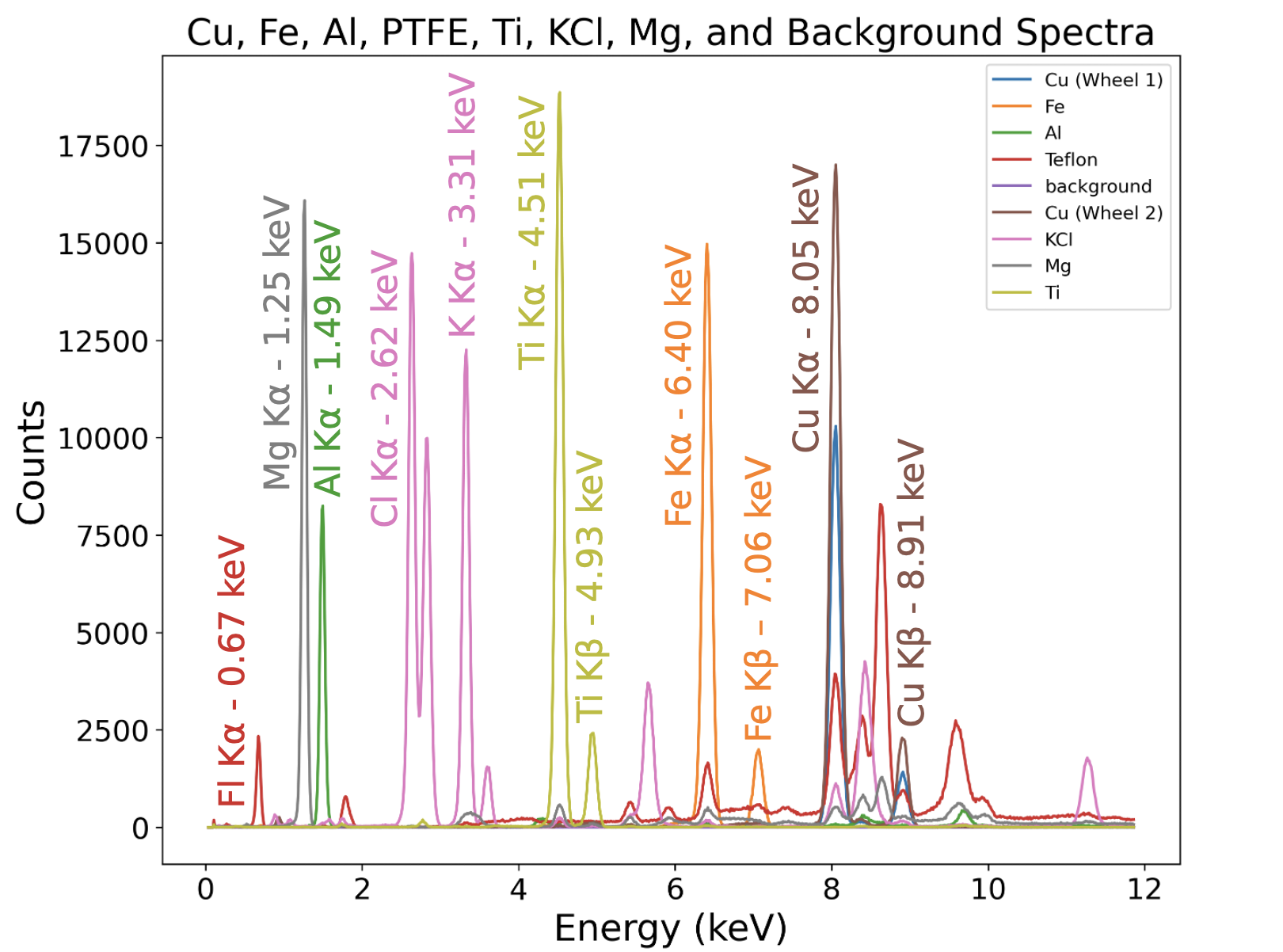}
   \end{tabular}
   \end{center}
   \caption[example] 
   {\label{fig:sddspectra} 
Superposed spectra for each of the seven beamline targets (teflon, magnesium, aluminum, KCl, titanium, iron, and copper) as measured by the SDD. Collection times varied from 60 seconds (copper) to 1450 seconds (teflon). Count rates ranged from 10 counts/second (teflon) to 2800 counts/second (copper).}
\end{figure}

In addition to the CCD spectra, in Fig. \ref{fig:sddspectra}, SDD spectra of all 7 targets across two target wheels (including an additional copper target that was placed on the second target wheel) were collected and are plotted against a background spectrum. SDD collection times varied from 60 seconds (copper) to 1450 seconds (teflon), and count rates of between 10 counts per second (teflon) and 2800 counts per second (copper) were recorded. It should be noted that the C2 window on the SDD attenuates low energy X-rays; roughly 50-60 \% of the 0.67 keV fluorine photons from the teflon target are transmitted, and roughly 60-70 \% of the 1.25 keV magnesium and 1.49 keV aluminum photons. It is therefore assumed that the true fluxes of low energy lines at the detector locations are $1.4-2\times$ greater than recorded by the SDD. The SDD spectra also reveal peaks from contamination sources, including the X-ray source tungsten anode L$\alpha$ line at 8.41 keV, as well as the characteristic K$\alpha$ and K$\beta$ lines from neighboring targets due to imperfect alignment of the target wheel. Other possible contamination sources which appeared in all spectra to varying degrees may be coming from the stainless steel of the beamline body or the aluminum 6061 alloy target wheel (silicon - 1.74 keV K$\alpha$, chromium - 5.41 keV K$\alpha$, manganese - 5.90 keV K$\alpha$, zinc - 8.64 keV K$\alpha$). To mitigate this contamination, in future XRF wheel interations we plan to use a pure aluminum wheel. 

\section{Summary and future plans}
\label{sec:summary}

With the XOC X-ray beamline fully assembled and commissioned, and all regulatory and monitoring systems and safety fail-safes working reliably, we are able to carry out X-ray detector characterization across a broad range of energies in the 0.3 - 10 keV band, under vacuum and down to temperatures as low as 173 K. The beamline has demonstrated reasonable count rates for teflon, aluminum, magnesium, potassium chloride, titanium, iron, and copper targets, with minimal contamination and radiation leakage observed. CCD spectra indicate that within the beamline, we approach the fano limited resolution of the detectors, especially for higher energy targets. With this robust new test system, the XOC group will soon move on to characterize new MIT-LL CCID-93++ AXIS prototype detectors. These detectors will feature both p-JFET outputs and multiple SiSeRO output variants. With multi-channel ASIC readout also enabled in the beamline, this test system will allow us to characterize and optimize the speed and noise performances of these new devices.

\acknowledgments 
 
We acknowledge support from the Kavli Institute for Particle Astrophysics and Cosmology at Stanford through a KIPAC Decadal Grant. We thank the Stanford Physics Machine shop for their assistance in modifying the detector chamber door to enable efficient detector and Archon controller mounting. Scientific measurements being made with the XOC beamline are supported by NASA APRA grants 80NSSC19K0499 and 80NSSC22K1921.

\begin{appendix}

\section{{\ttfamily GEANT4} Physics List Particle Interactions}
\label{sec:g4physics}

For gamma particles or photons:
\newline {\ttfamily G4PhotoElectricEffect}, {\ttfamily G4LivermorePhotoElectricModel}, {\ttfamily G4ComptonScattering}, \newline {\ttfamily G4LivermorePolarizedComptonModel}, {\ttfamily G4GammaConversion}, {\ttfamily G4RayleighScattering}, 
\newline {\ttfamily G4LivermorePolarizedRayleighModel}, {\ttfamily G4EmProcessOptions}, and {\ttfamily G4EmLivermorePhysics} 

\noindent For electrons: 
\newline {\ttfamily G4eMultipleScattering}, {\ttfamily G4eIonisation}, {\ttfamily \newline G4LivermoreIonisationModel}, {\ttfamily G4eBremsstrahlung}, {\ttfamily G4LivermoreBremsstrahlungModel}, {\ttfamily \newline G4eplusAnnihilation}, {\ttfamily G4UAtomicDeexcitation}, and {\ttfamily G4LossTableManager}.

\section{{\ttfamily GEANT4} Material Fractional Compositions}
\label{sec:materials}

\begin{table}[ht]
\caption{304L Stainless Steel}
\begin{center}       
\begin{tabular}{|c|c|} 
\hline
\rule[-1ex]{0pt}{3.5ex} Element & Mass Fraction \\
\hline
\rule[-1ex]{0pt}{3.5ex} Fe & 0.6931 \\
\hline
\rule[-1ex]{0pt}{3.5ex}  Cr & 0.185  \\
\hline
\rule[-1ex]{0pt}{3.5ex}  Ni & 0.09 \\
\hline
\rule[-1ex]{0pt}{3.5ex}  Mn & 0.02 \\
\hline
\rule[-1ex]{0pt}{3.5ex}  Si & 0.01 \\
\hline
\rule[-1ex]{0pt}{3.5ex}  N & 0.001 \\
\hline
\rule[-1ex]{0pt}{3.5ex}  P & 0.00045 \\
\hline
\rule[-1ex]{0pt}{3.5ex}  C & 0.0003 \\
\hline
\rule[-1ex]{0pt}{3.5ex}  S & 0.00015 \\
\hline
\end{tabular}
\end{center}
\end{table}

\begin{table}[ht]
\caption{PCB}
\begin{center}       
\begin{tabular}{|c|c|} 
\hline
\rule[-1ex]{0pt}{3.5ex} Element & Mass Fraction \\
\hline
\rule[-1ex]{0pt}{3.5ex} Al & 0.3904 \\
\hline
\rule[-1ex]{0pt}{3.5ex}  O & 0.3045  \\
\hline
\rule[-1ex]{0pt}{3.5ex}  Si & 0.1414 \\
\hline
\rule[-1ex]{0pt}{3.5ex}  Cu & 0.0739 \\
\hline
\rule[-1ex]{0pt}{3.5ex}  Sn & 0.0488 \\
\hline
\rule[-1ex]{0pt}{3.5ex}  C & 0.041 \\
\hline
\end{tabular}
\end{center}
\end{table}

\begin{table}[ht]
\caption{G10}
\begin{center}       
\begin{tabular}{|c|c|} 
\hline
\rule[-1ex]{0pt}{3.5ex} Element & Mass Fraction \\
\hline
\rule[-1ex]{0pt}{3.5ex}  O & 0.385764 \\
\hline
\rule[-1ex]{0pt}{3.5ex}  H & 0.275853  \\
\hline
\rule[-1ex]{0pt}{3.5ex}  Si & 0.151423 \\
\hline
\rule[-1ex]{0pt}{3.5ex}  Ca & 0.081496 \\
\hline
\rule[-1ex]{0pt}{3.5ex}  Al & 0.044453 \\
\hline
\rule[-1ex]{0pt}{3.5ex}  N & 0.027945 \\
\hline
\rule[-1ex]{0pt}{3.5ex}  B & 0.01864 \\
\hline
\rule[-1ex]{0pt}{3.5ex}  Mg & 0.010842 \\
\hline
\rule[-1ex]{0pt}{3.5ex}  C & 0.003583 \\
\hline
\end{tabular}
\end{center}
\end{table}

\begin{table}[ht]
\caption{Al-6061}
\begin{center}       
\begin{tabular}{|c|c|} 
\hline
\rule[-1ex]{0pt}{3.5ex} Element & Mass Fraction \\
\hline
\rule[-1ex]{0pt}{3.5ex}  Al & 0.96 \\
\hline
\rule[-1ex]{0pt}{3.5ex}  Mg & 0.012  \\
\hline
\rule[-1ex]{0pt}{3.5ex}  Si & 0.008 \\
\hline
\rule[-1ex]{0pt}{3.5ex}  Fe & 0.007 \\
\hline
\rule[-1ex]{0pt}{3.5ex}  Cu & 0.004 \\
\hline
\rule[-1ex]{0pt}{3.5ex}  Cr & 0.0035 \\
\hline
\rule[-1ex]{0pt}{3.5ex}  Zn & 0.0025 \\
\hline
\rule[-1ex]{0pt}{3.5ex}  Mn & 0.0015 \\
\hline
\rule[-1ex]{0pt}{3.5ex}  Ti & 0.0015 \\
\hline
\end{tabular}
\end{center}
\end{table}

\begin{table}[ht]
\caption{Ceramic}
\begin{center}       
\begin{tabular}{|c|c|} 
\hline
\rule[-1ex]{0pt}{3.5ex} Element & Mass Fraction \\
\hline
\rule[-1ex]{0pt}{3.5ex}  Al & 0.6581 \\
\hline
\rule[-1ex]{0pt}{3.5ex}  N & 0.3419  \\
\hline
\end{tabular}
\end{center}
\end{table}

\end{appendix}

\clearpage

\begin{thebibliography}{10}

\bibitem{Janesick01}
Janesick, J.~R.,  [{\em Scientific Charge-Coupled Devices}{\nolinebreak\hspace{0.1em}]}, SPIE Press, Bellingham, Washington (2001).

\bibitem{Lesser15_ccd}
Lesser, M., ``A summary of charge-coupled devices for astronomy,'' {\em Publications of the Astronomical Society of the Pacific}~{\bf 127}(957),  1097 (2015).

\bibitem{gruner02_ccd}
Gruner, S.~M., Tate, M.~W., and Eikenberry, E.~F., ``Charge-coupled device area x-ray detectors,'' {\em Review of Scientific Instruments}~{\bf 73}(8),  2815--2842 (2002).

\bibitem{Decadal23}
{National Academies of Sciences, Engineering, and Medicine},  [{\em {Pathways to Discovery in Astronomy and Astrophysics for the 2020s: Highlights of a Decadal Survey}}{\nolinebreak\hspace{0.1em}]}, The National Academies Press, Washington, DC (2023).

\bibitem{Ballet99}
{Ballet}, J., ``{Pile-up on X-ray CCD Instruments},'' in [{\em Astronomical Data Analysis Software and Systems X}{\nolinebreak\hspace{0.1em}]},  {Harnden}, F.~R., J., {Primini}, F.~A., and {Payne}, H.~E., eds., {\em Astronomical Society of the Pacific Conference Series} {\bf 238},  381 (Jan. 2001).

\bibitem{McCollough05}
{McCollough}, M.~L. and {Rots}, A.~H., ``{The Impact of the ACIS Readout Streak and Pileup on Chandra Source Detection},'' in [{\em Astronomical Data Analysis Software and Systems XIV}{\nolinebreak\hspace{0.1em}]},  {Shopbell}, P., {Britton}, M., and {Ebert}, R., eds., {\em Astronomical Society of the Pacific Conference Series} {\bf 347},  478 (Dec. 2005).

\bibitem{lumb00_pileup_xmm}
{Lumb}, D.~H., ``{Simulations and Mitigation of Pile-Up in XMM CCD Instruments},'' {\em Experimental Astronomy}~{\bf 10},  439--456 (Nov. 2000).

\bibitem{HCMOS07}
{Bai}, Y., {Bajaj}, J., {Beletic}, J.~W., {Farris}, M.~C., {Joshi}, A., {Lauxtermann}, S., {Petersen}, A., and {Williams}, G., ``{Teledyne Imaging Sensors: Silicon CMOS imaging technologies for x-ray, UV, visible, and near infrared},'' in [{\em High Energy, Optical, and Infrared Detectors for Astronomy III}{\nolinebreak\hspace{0.1em}]},  {Dorn}, D.~A. and {Holland}, A.~D., eds., {\em Society of Photo-Optical Instrumentation Engineers (SPIE) Conference Series} {\bf 7021},  702102 (July 2008).

\bibitem{HCMOS17}
Hull, S.~V., Falcone, A.~D., Burrows, D.~N., Wages, M., Chattopadhyay, T., McQuaide, M., Bray, E., and Kern, M., ``{Recent X-ray hybrid CMOS detector developments and measurements},'' in [{\em UV, X-Ray, and Gamma-Ray Space Instrumentation for Astronomy XX}{\nolinebreak\hspace{0.1em}]},  Siegmund, O.~H., ed.,  {\bf 10397},  1039704, International Society for Optics and Photonics, SPIE (2017).

\bibitem{DEPFET20}
{Treberspurg}, W., {Andritschke}, R., {Behrens}, A., {Bonholzer}, M., {Emberger}, V., {Hauser}, G., {Lechner}, P., {Meidinger}, N., and {M{\"u}ller-Seidlitz}, J., ``{Characterization of a 256 {\texttimes} 256 pixel DEPFET detector for the WFI of Athena},'' {\em Nuclear Instruments and Methods in Physics Research A}~{\bf 958},  162555 (Apr. 2020).

\bibitem{chattopadhyay18_HCDoverview}
Chattopadhyay, T., Falcone, A.~D., Burrows, D.~N., Hull, S., Bray, E., Wages, M., McQuaide, M., Buntic, L., Crum, R., O'Dell, J., and Anderson, T., ``X-ray hybrid cmos detectors: Recent development and characterization progress,'' in [{\em Space Telescopes and Instrumentation 2018: Ultraviolet to Gamma Ray}{\nolinebreak\hspace{0.1em}]},  {\em Proc.SPIE}~{\bf 10709} (2018).

\bibitem{andricekDEPFET22}
Andricek, L., Bähr, A., Lechner, P., Ninkovic, J., Richter, R., Schopper, F., and Treis, J., ``Depfet—recent developments and future prospects,'' {\em Frontiers in Physics}~{\bf 10} (06 2022).

\bibitem{chattopadhyay20}
Chattopadhyay, T., Herrmann, S., Allen, S.~W., Hirschman, J., Morris, G., Bautz, M., Malonis, A., Foster, R., Prigozhin, G., Craig, D., and Burke, B., ``{Tiny-box: a tool for the versatile development and characterization of low noise fast x-ray imaging detectors},'' in [{\em X-Ray, Optical, and Infrared Detectors for Astronomy IX}{\nolinebreak\hspace{0.1em}]},   {\bf 11454},  1145423, International Society for Optics and Photonics, SPIE (2020).

\bibitem{BeWindow}
{Brackney}, H. and {Atlee}, Z.~J., ``{Beryllium Windows for Permanently Evacuated X-Ray Tubes},'' {\em Review of Scientific Instruments}~{\bf 14},  59--63 (Mar. 1943).

\bibitem{LOWE1997354}
Lowe, B., ``Measurements of fano factors in silicon and germanium in the low-energy x-ray region,'' {\em Nuclear Instruments and Methods in Physics Research Section A: Accelerators, Spectrometers, Detectors and Associated Equipment}~{\bf 399}(2),  354--364 (1997).

\bibitem{WangPID2018}
Wang, Y.,  [{\em PID Temperature Control}{\nolinebreak\hspace{0.1em}]},  63--76, Springer International Publishing, Cham (2018).

\bibitem{SDD}
Vacchi, A.,  [{\em {Silicon Drift Detectors. In: Bambi, C., Santangelo, A. (eds) Handbook of X-ray and Gamma-ray Astrophysics}}{\nolinebreak\hspace{0.1em}]}, Springer, Singapore (2023).

\bibitem{bautz18}
Bautz, M., Foster, R., LaMarr, B., Malonis, A., Prigozhin, G., Miller, E., Grant, C.~E., Burke, B., Cooper, M., Craig, D., Leitz, C., Schuette, D., and Suntharalingam, V., ``{Toward fast low-noise low-power digital CCDs for Lynx and other high-energy astrophysics missions},'' in [{\em Space Telescopes and Instrumentation 2018: Ultraviolet to Gamma Ray}{\nolinebreak\hspace{0.1em}]},  den Herder, J.-W.~A., Nikzad, S., and Nakazawa, K., eds.,  {\bf 10699},  238 -- 248, International Society for Optics and Photonics, SPIE (2018).

\bibitem{bautz19}
{Bautz}, M.~W., {Burke}, B.~E., {Cooper}, M., {Craig}, D., {Foster}, R.~F., {Grant}, C.~E., {LaMarr}, B.~J., {Leitz}, C., {Malonis}, A., {Miller}, E.~D., {Prigozhin}, G., {Schuette}, D., {Suntharalingam}, V., and {Thayer}, C., ``{Toward fast, low-noise charge-coupled devices for Lynx},'' {\em Journal of Astronomical Telescopes, Instruments, and Systems}~{\bf 5},  021015 (Apr. 2019).

\bibitem{bautz20}
Bautz, M., Burke, B., Cooper, M., Craig, D., Donlon, K., Foster, R., Grant, C.~E., LaMarr, B., Leitz, C., Malonis, A., Miller, E., Prigozhin, G., Thayer, C., Allen, S., Herrmann, S., Chattopadhyay, T., and Morris, R.~G., ``{Progress toward fast, low-noise, low-power CCDs for Lynx and other high-energy astrophysics missions},'' in [{\em Space Telescopes and Instrumentation 2020: Ultraviolet to Gamma Ray}{\nolinebreak\hspace{0.1em}]},  den Herder, J.-W.~A., Nikzad, S., and Nakazawa, K., eds.,  {\bf 11444},  1144494, International Society for Optics and Photonics, SPIE (2020).

\bibitem{bautz22}
Bautz, M., Foster, R., Grant, C.~E., LaMarr, B., Malonis, A., Miller, E., Prigozhin, G., Burke, B., Cooper, M., Donlon, K., Lambert, R., Warner, K., Young, D., Chattopadhyay, T., Herrmann, S., Morris, R.~G., Leitz, C., and Allen, S., ``{Performance of high frame-rate x-ray CCDs for future strategic missions},'' in [{\em Space Telescopes and Instrumentation 2022: Ultraviolet to Gamma Ray}{\nolinebreak\hspace{0.1em}]},  den Herder, J.-W.~A., Nikzad, S., and Nakazawa, K., eds.,  {\bf 12181},  121812A, International Society for Optics and Photonics, SPIE (2022).

\bibitem{tanmoyJATIS22}
{Chattopadhyay}, T., {Herrmann}, S., {Orel}, P., {Morris}, R.~G., {Prigozhin}, G., {Malonis}, A., {Foster}, R., {Craig}, D., {Burke}, B.~E., {Allen}, S.~W., and {Bautz}, M.~W., ``{Development and characterization of a fast and low noise readout for the next generation x-ray charge-coupled devices},'' {\em Journal of Astronomical Telescopes, Instruments, and Systems}~{\bf 8},  026005 (Apr. 2022).

\bibitem{herrmann20_mcrc}
Herrmann, S., Wong, J., Chattopadhyay, T., Morris, R.~G., Burke, B., Prigozhin, G., Cooper, M., Craig, D., Donlon, K., Foster, R., Malonis, A., Bautz, M., and Allen, S., ``{MCRC V1: development of integrated readout electronics for next generation x-ray CCD detectors for future satellite observatories},'' in [{\em X-Ray, Optical, and Infrared Detectors for Astronomy IX}{\nolinebreak\hspace{0.1em}]},  Holland, A.~D. and Beletic, J., eds.,  {\bf 11454},  412 -- 418, International Society for Optics and Photonics, SPIE (2020).

\bibitem{porelMCRCspie2022}
Orel, P., Herrmann, S., Chattopadhyay, T., Morris, G.~R., Allen, S.~W., Prigozhin, G.~Y., Foster, R., Malonis, A., Bautz, M.~W., Cooper, M.~J., and Donlon, K., ``{X-ray speed reading with the MCRC: a low noise CCD readout ASIC enabling readout speeds of 5 Mpixel/s/channel},'' in [{\em X-Ray, Optical, and Infrared Detectors for Astronomy X}{\nolinebreak\hspace{0.1em}]},  Holland, A.~D. and Beletic, J., eds.,  {\bf 12191},  1219124, International Society for Optics and Photonics, SPIE (2022).

\bibitem{tanmoyspie2024}
Chattopadhyay, T., Herrmann, S.~C., Orel, P., Donlon, K., Allen, S.~W., Bautz, M.~W., Cantrall, B.~J., Cooper, M.~J., LaMarr, B.~J., Leitz, C.~W., Miller, E.~D., Morris, G.~R., Prigozhin, G.~Y., Prigozhin, I., Stueber, H.~R., and Wilkins, D.~R., ``{Demonstrating sub-electron noise performance in Single electron Sensitive Readout (SiSeRO) devices},'' in [{\em X-Ray, Optical, and Infrared Detectors for Astronomy XI}{\nolinebreak\hspace{0.1em}]},   {\bf 13103},  1310357, International Society for Optics and Photonics, SPIE (2024).

\bibitem{bevspie2024}
LaMarr, B.~J., Schneider, B., Prigozhin, G.~Y., Miller, E.~D., Bautz, M.~W., Foster, R.~F., Grant, C.~E., Malonis, A.~C., Cooper, M.~J., Lambert, R.~D., Ryu, K.~K., and Jensen, M., ``{Soft X-ray resolution and scientific performance of CCD sensors for future X-ray missions},'' in [{\em X-Ray, Optical, and Infrared Detectors for Astronomy XI}{\nolinebreak\hspace{0.1em}]},   {\bf 13103},  1310333, International Society for Optics and Photonics, SPIE (2024).

\bibitem{svenspie2024}
Herrmann, S.~C., Orel, P., Chattopadhyay, T., Morris, G.~R., Prigozhin, G.~Y., Stueber, H.~R., Allen, S.~W., Bautz, M.~W., Donlon, K., LaMarr, B.~J., Leitz, C.~W., Miller, E.~D., Poliszczuk, A., and Wilkins, D.~R., ``{Continued developments in X-ray speed reading: fast, low noise readout for next-generation wide-field imagers},'' in [{\em X-Ray, Optical, and Infrared Detectors for Astronomy XI}{\nolinebreak\hspace{0.1em}]},   {\bf 13103},  1310383, International Society for Optics and Photonics, SPIE (2024).

\bibitem{porelMCRCspie2024}
Orel, P., Herrmann, S., Chattopadhyay, T., Morris, G.~R., Stueber, H., Pan, A., Allen, S.~W., Wilkins, D., Prigozhin, G.~Y., LaMarr, B., Foster, R., Malonis, A., Bautz, M.~W., Cooper, M.~J., and Donlon, K., ``{X-ray speed reading with the MCRC: prototype success and next generation upgrades},'' in [{\em X-Ray, Optical, and Infrared Detectors for Astronomy XI}{\nolinebreak\hspace{0.1em}]},   {\bf 13103},  1310332, International Society for Optics and Photonics, SPIE (2024).

\bibitem{sisero2021}
{Chattopadhyay}, T., {Herrmann}, S., {Burke}, B.~E., {Donlon}, K., {Prigozhin}, G., {Morris}, G., {Orel}, P., {Cooper}, M., {Malonis}, A., {Wilkins}, D.~R., {Suntharalingam}, V., {Allen}, S.~W., {Bautz}, M.~W., and {Leitz}, C., ``{First results on SiSeRO devices: a new x-ray detector for scientific instrumentation},'' {\em Journal of Astronomical Telescopes, Instruments, and Systems}~{\bf 8},  026006 (Apr. 2022).

\bibitem{sisero2022}
{Chattopadhyay}, T., {Herrmann}, S., {Orel}, P., {Morris}, R.~G., {Wilkins}, D.~R., {Allen}, S.~W., {Prigozhin}, G., {LaMarr}, B., {Malonis}, A., {Foster}, R., {Bautz}, M.~W., {Donlon}, K., {Cooper}, M., and {Leitz}, C., ``{Single electron sensitive readout (SiSeRO) x-ray detectors: technological progress and characterization},'' in [{\em X-Ray, Optical, and Infrared Detectors for Astronomy X}{\nolinebreak\hspace{0.1em}]},  {Holland}, A.~D. and {Beletic}, J., eds., {\em Society of Photo-Optical Instrumentation Engineers (SPIE) Conference Series} {\bf 12191},  121910W (Aug. 2022).

\bibitem{sisero2023}
{Chattopadhyay}, T., {Herrmann}, S., {Kaplan}, M., {Orel}, P., {Donlon}, K., {Prigozhin}, G., {Morris}, G., {Cooper}, M., {Malonis}, A., {Allen}, S.~W., {Bautz}, M.~W., and {Leitz}, C., ``{Improved noise performance from the next-generation buried-channel p-MOSFET SiSeROs},'' {\em Journal of Astronomical Telescopes, Instruments, and Systems}~{\bf 9},  026001 (Apr. 2023).

\bibitem{MCRC22}
Chattopadhyay, T., Herrmann, S., Orel, P., Morris, G., Prigozhin, G., Malonis, A., Foster, R., Craig, D., Burke, B., Allen, S., and Bautz, M., ``Development and characterization of a fast and low noise readout for the next generation x-ray charge-coupled devices.,'' {\em Journal of Astronomical Telescopes, Instruments, and Systems}~{\bf 8},  1--12 (2022).

\bibitem{AGOSTINELLI2003}
Agostinelli, S., Allison, J., Amako, K., Apostolakis, J., Araujo, H., Arce, P., Asai, M., Axen, D., Banerjee, S., Barrand, G., Behner, F., Bellagamba, L., Boudreau, J., Broglia, L., Brunengo, A., Burkhardt, H., Chauvie, S., Chuma, J., Chytracek, R., Cooperman, G., Cosmo, G., Degtyarenko, P., Dell'Acqua, A., Depaola, G., Dietrich, D., Enami, R., Feliciello, A., Ferguson, C., Fesefeldt, H., Folger, G., Foppiano, F., Forti, A., Garelli, S., Giani, S., Giannitrapani, R., Gibin, D., {Gómez Cadenas}, J., González, I., {Gracia Abril}, G., Greeniaus, G., Greiner, W., Grichine, V., Grossheim, A., Guatelli, S., Gumplinger, P., Hamatsu, R., Hashimoto, K., Hasui, H., Heikkinen, A., Howard, A., Ivanchenko, V., Johnson, A., Jones, F., Kallenbach, J., Kanaya, N., Kawabata, M., Kawabata, Y., Kawaguti, M., Kelner, S., Kent, P., Kimura, A., Kodama, T., Kokoulin, R., Kossov, M., Kurashige, H., Lamanna, E., Lampén, T., Lara, V., Lefebure, V., Lei, F., Liendl, M., Lockman, W., Longo, F., Magni, S., Maire, M., Medernach, E.,
  Minamimoto, K., {Mora de Freitas}, P., Morita, Y., Murakami, K., Nagamatu, M., Nartallo, R., Nieminen, P., Nishimura, T., Ohtsubo, K., Okamura, M., O'Neale, S., Oohata, Y., Paech, K., Perl, J., Pfeiffer, A., Pia, M., Ranjard, F., Rybin, A., Sadilov, S., {Di Salvo}, E., Santin, G., Sasaki, T., Savvas, N., Sawada, Y., Scherer, S., Sei, S., Sirotenko, V., Smith, D., Starkov, N., Stoecker, H., Sulkimo, J., Takahata, M., Tanaka, S., Tcherniaev, E., {Safai Tehrani}, E., Tropeano, M., Truscott, P., Uno, H., Urban, L., Urban, P., Verderi, M., Walkden, A., Wander, W., Weber, H., Wellisch, J., Wenaus, T., Williams, D., Wright, D., Yamada, T., Yoshida, H., and Zschiesche, D., ``Geant4—a simulation toolkit,'' {\em Nuclear Instruments and Methods in Physics Research Section A: Accelerators, Spectrometers, Detectors and Associated Equipment}~{\bf 506}(3),  250--303 (2003).

\bibitem{poole11}
{Poole}, C.~M., {Cornelius}, I., {Trapp}, J.~V., and {Langton}, C.~M., ``{A CAD Interface for GEANT4},'' {\em arXiv e-prints} ,  arXiv:1105.0963 (May 2011).

\end{thebibliography}

\end{document}